\pdfoutput=1

\documentclass[11pt]{article}

\usepackage[]{acl}
\usepackage{times}
\usepackage{latexsym}
\usepackage{amsmath}
\usepackage{amssymb}
\usepackage{orcidlink}
\usepackage{graphicx}
\usepackage{multirow}
\usepackage{pgf-pie}   
\usepackage{xcolor} 
\usepackage{tablefootnote}
\usepackage{tabularx}
\usepackage{array}
\usepackage{seqsplit}

\newcolumntype{P}[1]{>{\raggedright\arraybackslash}p{#1}}
\newcolumntype{Y}{>{\raggedright\arraybackslash}X}
\newcolumntype{S}[1]{>{\raggedright\arraybackslash}p{#1}}
\newcommand{\schematable}[1]{%
  \setlength{\tabcolsep}{4pt}%
  \begin{tabularx}{\columnwidth}{@{}S{2.5cm}S{1.55cm}Y@{}}#1\end{tabularx}%
}
\usepackage{adjustbox}
\usepackage{float}
\usepackage{graphicx}
\usepackage{caption}
\usepackage{subcaption}
\usepackage{threeparttable}
\usepackage{flushend}
\usepackage{longtable}

\usepackage{lipsum}

\usepackage{booktabs}
\usepackage{colortbl}
\usepackage{algorithm}
\usepackage{algpseudocode}
\usepackage{tikz}
\usetikzlibrary{positioning,arrows.meta,fit,backgrounds,calc,shapes.geometric}
\definecolor{vihblue}{HTML}{2C6FBB}
\definecolor{vihaccent}{HTML}{E08A1E}
\definecolor{vihgray}{HTML}{6B6B6B}
\definecolor{vihbg}{HTML}{EAF1FB}

\usepackage{xcolor}
\usepackage{soul}

\usepackage[T5]{fontenc}

\usepackage[protrusion=false,nopatch=showhyphens]{microtype}

\title{ViHoRec: A Quality-Controlled Vietnamese Hotel Recommendation Dataset and Short-History Benchmark}

\author{
  Minh Hoang Nguyen\textsuperscript{1}\,\orcidlink{0009-0004-1384-3856}
  \and Tin Van Huynh\textsuperscript{2}
  \and Kiet Van Nguyen\textsuperscript{2}\\[0.4em]
  \textsuperscript{1}University of Science, Ho Chi Minh City, Vietnam\\
  \textsuperscript{2}University of Information Technology, Ho Chi Minh City, Vietnam\\
  Vietnam National University, Ho Chi Minh City, Vietnam\\[0.3em]
  {\tt 24C15049@student.hcmus.edu.vn} \quad
  {\tt tinhv@uit.edu.vn} \quad
  {\tt kietnv@uit.edu.vn}
}

\begin{document}
\maketitle
\begin{abstract}

Public recommendation resources for Vietnamese are scarce, and hotel data add two construction problems: one property is listed under different names on different booking platforms, and reviewers are exposed only through short display names. We release \textbf{ViHoRec}, 17{,}911 interactions between 6{,}822 reviewer keys and 560 hotels in nine Vietnamese destinations, collected from Booking.com, Traveloka, and Ivivu, with content metadata for 309 hotels. The pipeline resolves listings with a city- and type-aware key, audits quality with reproducible checks, and replaces display names with HMAC-SHA256 pseudonyms. On 426 stratified candidate pairs the released resolver makes no false merge and misses one of 22 duplicates, and correcting the miss leaves the benchmark unchanged. Because a key is a display name rather than a person, the short-history benchmark is reported on the released 798-key split and repeated on a stricter 481-key pool. All comparable models share one protocol: training history only at test time, the full 535-hotel catalogue, training items masked, and validation used only for model selection. SASRec reaches Recall@10 of $0.168 \pm 0.006$ over five seeds against 0.129 for UserKNN; on the stricter pool the two are not significantly different. A history-gated fusion of UserKNN and content is released as a diagnostic baseline; its differences from UserKNN are not significant globally or in any history bucket. Available dataset at \url{https://huggingface.co/datasets/MinhDS/ViHoRec}.


\end{abstract}
\section{Introduction}
\label{sec:introduction}

\begin{figure*}[!t]
    \centering
    \includegraphics[width=\textwidth]{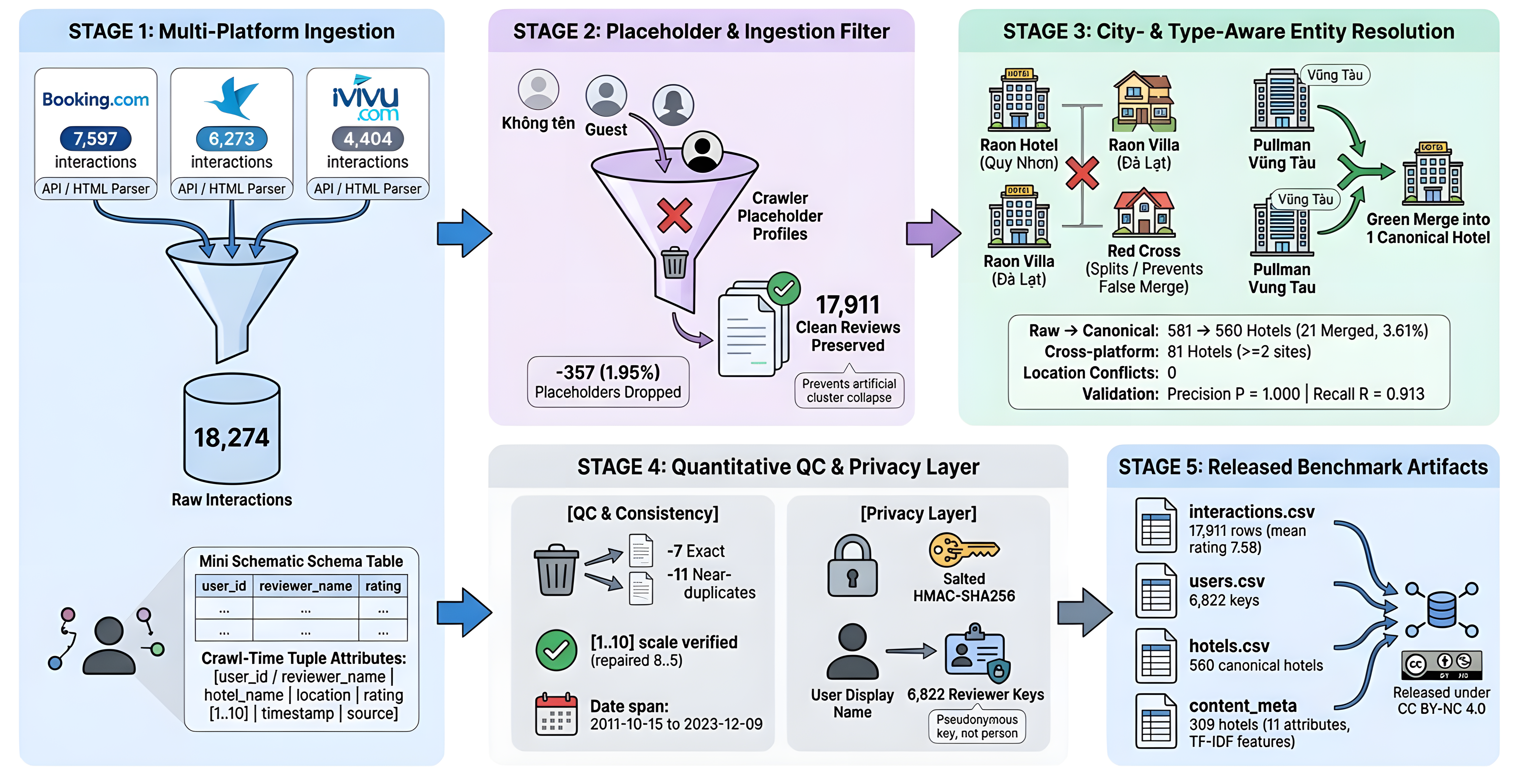}
    \caption{Overview of \textbf{ViHoRec}. Public reviews from three booking platforms are crawled, resolved into canonical hotel entities across sites, quality-controlled, and \emph{pseudonymized} (HMAC of display names; identifiers are not anonymized), then released as three tables (plus content metadata) together with a reproducible public split and baseline benchmark.}
    \label{fig:overview}
\end{figure*}

Hotel recommendation is a central task in online travel platforms: given a user's past bookings or ratings, the system must rank accommodations that match traveler preferences under sparse and noisy interaction logs~\cite{chaudhari2020travel,antognini2020hotelrec,chen2023travelkw}. Progress on this task is tightly coupled to the availability of public interaction resources with documented construction quality~\cite{harper2015movielens,gebru2021datasheets}. This paper focuses on the specific setting of Vietnamese hotel recommendation, where the input is multi-platform reviewer--hotel ratings and hotel content metadata, and the output is a ranked list under a public short-history evaluation protocol.

Vietnamese remains data-poor for this setting. Widely used benchmarks (MovieLens, Amazon Reviews, Yelp) and large hotel corpora such as HotelRec are English-centric~\cite{harper2015movielens,he2016ups,antognini2020hotelrec}, so models trained on them do not transfer language, platform conventions, or market sparsity. To the best of our knowledge, no public Vietnamese hotel recommendation dataset jointly documents collection, cross-platform entity resolution, quantitative quality control, and a short-history ranking protocol. Constructing such a resource is challenging for three reasons. \textbf{First}, the same hotel appears on Booking.com, Traveloka, and Ivivu under inconsistent names, so interactions are not comparable without explicit entity resolution~\cite{kopcke2012entity}. \textbf{Second}, scraped ratings require auditable cleaning: duplicates, missing fields, placeholders, and cross-site inconsistencies must be measured rather than assumed~\cite{gebru2021datasheets}. \textbf{Third}, a public release must remove display names yet remain benchmarkable. The platforms expose only short display names, so the only available identifier is a key derived from the name; it is not a person, and two travellers who share a name share a key. Most keys in the corpus have a single interaction, and a leave-one-out ranking protocol can evaluate only keys that still have training history after the holdout.

Methods for sparse and cold-start recommendation have advanced quickly, from multi-criteria hotel models on TripAdvisor-style corpora to meta-learning and attribute-graph encoders that adapt from a few interactions~\cite{nilashi2015multicriteria,jannach2014multicriteria,lee2019melu,qian2020attribute,lin2024fanp,weimann2025felrec}. All of them are validated on English or multi-domain benchmarks with denser side information than a low-resource hotel market provides, so their value for Vietnamese hotels cannot be measured until a public, quality-controlled testbed exists.

We address this resource gap with \textbf{ViHoRec} (Figure~\ref{fig:overview}), a quality-controlled Vietnamese hotel recommendation dataset and short-history benchmark crawled from Booking.com, Traveloka, and Ivivu. The primary contribution is the construction itself: a documented pipeline that makes sparse Vietnamese hotel interactions comparable across platforms and evaluable under one fixed public protocol. In contrast to English hotel corpora that emphasize scale, ViHoRec emphasizes auditability, and it reports the limits of its own identifiers and entity links as measured quantities. Our contributions are:
\begin{enumerate}
    \item \textbf{ViHoRec Dataset.} We release 17{,}911 cleaned interactions between 6{,}822 reviewer keys and 560 hotels in nine destinations, with content metadata for 309 hotels. Hotel listings are reconciled across platforms by a city- and type-aware resolver whose precision, sample recall, and false-merge and false-split rates are reported with exact confidence intervals on 426 stratified pairs, and whose one confirmed miss and one disputed pair are propagated to the benchmark.
    \item \textbf{Privacy-preserving public release.} Display names are replaced by HMAC-SHA256 pseudonyms (pseudonymization, not anonymization). A pseudonym is a \emph{reviewer key}, not a person: we measure a lower bound on key collisions and repeat the benchmark on a stricter key pool.
    \item \textbf{Short-history benchmark.} We publish a three-way temporal leave-one-out split, chronological within each reviewer key, with validation-only model selection. Every comparable model uses training history only at test time, ranks the full catalogue, and masks training items; results carry bootstrap intervals and paired tests, and numbers obtained under any other setting are kept in a separate reference table.
    \item \textbf{A diagnostic baseline.} Adaptive Hybrid gates UserKNN and Content-TFIDF by history length. It is included to show how the released metadata behaves under short histories, not as a recommended ranker: its differences from UserKNN are not significant globally or in any history bucket.
\end{enumerate}

The remainder of this paper is organized as follows. Section~\ref{sec:related} reviews hotel recommenders, cold-start methods, and public datasets. Section~\ref{sec:dataset} presents the dataset: collection and entity resolution, quality control and validation (protocol in Appendix~\ref{app:guidelines}), and the released files and their statistics (full schemas in Appendix~\ref{app:schema}). Section~\ref{sec:methods} defines the diagnostic Adaptive Hybrid baseline. Section~\ref{sec:experiments} specifies the protocol, metrics, and baselines, reports the comparable results and the reference numbers, and analyzes reviewer-key sensitivity, entity-resolution sensitivity, history length, and the hybrid gate (earlier two-way ablations in Appendix~\ref{app:ablations}). Section~\ref{sec:ethics} discusses pseudonymization, ethics, and licensing, and Section~\ref{sec:conclusion} concludes with limitations and future work.

\section{Related Work}
\label{sec:related}

\begin{table*}[!ht]
\centering
\caption{ViHoRec versus representative recommendation and tourism datasets, plus two protocol-focused benchmarks discussed in Section~\ref{sec:related}. \textit{Meta.}: side information beyond scalar ratings (MC = multi-criteria; dest.\ = destination context; attrs = structured attributes); \textit{X-plat.}: explicit cross-platform hotel entity resolution; \textit{QC}: published construction / quality-control protocol for recommendation use. Interaction counts follow the cited releases.}
\label{tab:datasetcmp}
\footnotesize
\setlength{\tabcolsep}{4pt}
\begin{tabular}{@{}l l r c c c c@{}}
\toprule
\textbf{Dataset} & \textbf{Domain} & \textbf{\#Inter.} & \textbf{Lang.} & \textbf{Meta.} & \textbf{X-plat.} & \textbf{QC} \\
\midrule
\multicolumn{7}{@{}l}{\textit{General recommender benchmarks}} \\
MovieLens-100K~\cite{harper2015movielens} & Movies & 100{,}000 & EN & genres & -- & \checkmark \\
Amazon Reviews~\cite{mcauley2015image,he2016ups} & E-comm. & $\sim$142M & EN & text & -- & \checkmark \\
Yelp Open~\cite{yelp2023opendataset} & Local biz. & $\sim$7.0M & EN & attrs & -- & \checkmark \\
\midrule
\multicolumn{7}{@{}l}{\textit{Dataset-focused / protocol benchmarks}} \\
KuaiRec~\cite{gao2022kuairec} & Short video & $\sim$4.7M & ZH & attrs & -- & \checkmark \\
Tenrec~\cite{yuan2022tenrec} & Feeds & $\sim$140M & ZH & attrs & -- & \checkmark \\
\midrule
\multicolumn{7}{@{}l}{\textit{Hotel / travel resources}} \\
TripAdvisor MC~\cite{jannach2014multicriteria} & Hotels & $\sim$22{,}000 & EN & MC & -- & \checkmark \\
HotelRec~\cite{antognini2020hotelrec} & Hotels & 50M & EN & text & -- & \checkmark \\
Booking.com Trips~\cite{goldenberg2021booking} & Travel & 1.5M & multi & dest. & -- & \checkmark \\
Booking.com Reviews~\cite{igeharia2024bookingreviews} & Hotels & $\sim$2.0M & EN & attrs & -- & \checkmark \\
\midrule
\multicolumn{7}{@{}l}{\textit{Vietnamese tourism text}} \\
VN Hotel CX~\cite{nguyen2023understanding} & Hotels & $\sim$20{,}600 & VI & text & -- & -- \\
\rowcolor{vihbg}
\textbf{ViHoRec (ours)} & Hotels & 17{,}911 & \textbf{VI} & attrs & \checkmark & \checkmark \\
\bottomrule
\end{tabular}
\end{table*}

\subsection{Hotel and travel recommender systems}
A hotel recommender ranks properties for a traveller from past stays or ratings, usually with few interactions per traveller and strong destination effects~\cite{chaudhari2020travel}. Early hotel systems apply multi-criteria CF, bipartite link prediction, or location-aware neighborhood ranking on TripAdvisor-style corpora~\cite{nilashi2015multicriteria,shambour2022hotel,jannach2014multicriteria,kaya2020hotel,alghobari2021lapta,levi2012needle}. More recent travel recommenders incorporate review keywords, geo-tagged visual similarity, and reinforcement learning over candidate attractions~\cite{chen2023travelkw,chen2023tourist}, while next-POI models mine geographical and interactive correlations under sparse mobility logs~\cite{ren2023mining} and sequential models forecast hotel reservations from booking histories~\cite{wang2020forecasting}. Hybrid designs typically fuse collaborative and content signals through fixed late fusion, feature concatenation, or ontology-based similarity~\cite{burke2002hybrid,kuznetsov2025improving}. Across this line of work, two assumptions recur: an English or high-resource interaction corpus already exists, and side information is dense enough to compensate for short user histories. Neither assumption holds for Vietnamese multi-platform hotel data, where the missing piece is a public, quality-controlled interaction resource rather than another proprietary model.

Point-of-interest (POI) recommendation on location-based social networks is the closest adjacent line. It models check-in sequences with time-aware, cross-domain, and spatial-graph signals~\cite{acharya2025timeaware,acharya2024highorder}, and a recent survey traces how these signals have been combined~\cite{acharya2025poisurvey}. ViHoRec addresses a different task: it ranks hotels from review interactions pooled across three booking platforms, so its construction problems are cross-platform entity resolution, display-name identity, and a public short-history split, rather than next-check-in prediction from geo-tagged social data. Because the task and the inputs differ, these methods are discussed as adjacent work and are not run as baselines on this dataset.

\subsection{Cold-start recommendation and hybrid fusion}
Cold-start remains a core obstacle whenever users or items lack sufficient interactions~\cite{schein2002cold}. Recent methods attack this problem with meta-learning and representation transfer: MeLU and MetaHIN adapt quickly from few interactions~\cite{lee2019melu,lu2020meta}; MetaCF accelerates cold-start collaborative filtering~\cite{wei2020fast}; MetaKG and FANP combine knowledge graphs or neural processes with few-shot adaptation~\cite{du2023metakg,lin2024fanp}; and AGNN targets strict cold start via attribute graphs rather than interaction graphs~\cite{qian2020attribute}. Complementary work studies cross-domain latent feature mapping~\cite{wang2020cdlfm}, hierarchical GNN fusion of auxiliary information~\cite{zhang2024hi}, and dynamic sequential representations that update new items without side information or gradient-based finetuning~\cite{weimann2025felrec}. News recommendation uses the same long- and short-term split in a different domain: Co-NAML-LSTUR joins attentive multi-view content encoding with an LSTUR user representation~\cite{nguyen2025co}. Neighborhood CF remains sensitive to the choice of user-similarity metric on sparse datasets~\cite{sgardelis2025evaluation}; ontology-aware hybrids further show that content can remain usable from cold-start through denser regimes~\cite{kuznetsov2025improving}. Soft-cluster side-information embeddings likewise stabilize item cold start without heavy neural stacks~\cite{puthiya2020simple,volkovs2017dropoutnet}. These methods set the comparative baseline ceiling that any new resource should support.

\subsection{Public datasets, entity resolution, and quality control}
Public interaction resources fall into three groups relevant to our setting. \emph{General recommender benchmarks}---MovieLens, Amazon Reviews, and Yelp---dominate offline CF evaluation at large English-centric scale~\cite{harper2015movielens,mcauley2015image,he2016ups,yelp2023opendataset,beel2019pruning}. \emph{Hotel and travel releases} include TripAdvisor multi-criteria subsets~\cite{jannach2014multicriteria}, the 50M-review HotelRec corpus~\cite{antognini2020hotelrec}, Booking.com multi-destination trips for sequential destination prediction~\cite{goldenberg2021booking}, and the Booking.com accommodation-review collection for review ranking~\cite{igeharia2024bookingreviews}. \emph{Vietnamese tourism text resources} release hotel reviews for customer-experience or sentiment analysis rather than recommendation splits~\cite{nguyen2023understanding,nguyen2024rrs}. Well-adopted datasets document collection and cleaning so that results are comparable~\cite{harper2015movielens,gebru2021datasheets}; when the same property is listed on multiple platforms, entity resolution is a prerequisite for coherent supervision~\cite{kopcke2012entity}.

Table~\ref{tab:datasetcmp} positions ViHoRec against these resources. Scale favors English hotel and e-commerce corpora, but none jointly provide Vietnamese hotel interactions, explicit cross-platform entity resolution, quantitative quality-control metrics, and a public short-history recommendation protocol. Booking.com trips and reviews are industry-grade yet target destination sequences or review ranking in English rather than Vietnamese multi-OTA hotel CF; Vietnamese CX releases share the language but not the recommendation task. To the best of our knowledge, ViHoRec is the first public Vietnamese hotel recommendation resource that fills this intersection.

\subsection{Dataset-focused benchmarks and evaluation protocols}
Recent recommendation resources are judged as much by their documentation and protocol as by their size. Table~\ref{tab:datasetcmp} therefore also lists two protocol-focused releases. KuaiRec releases a fully observed user--item matrix so that the bias of sparse offline logs can be measured~\cite{gao2022kuairec}, and Tenrec bundles several tasks over one platform with a fixed evaluation suite~\cite{yuan2022tenrec}. Reproducibility studies show that weak or untuned baselines, sampled metrics, and inconsistent splits can reverse reported rankings~\cite{dacrema2019progress,sun2020rigorous,krichene2020sampled}. Data splitting matters in particular: random and per-user leave-one-out splits differ from a global temporal cutoff, and per-user splits let information from a later period reach the training data of other users~\cite{meng2020splitting,ji2023leakage}. ViHoRec follows these lessons at small scale and in a low-resource language. It ranks the full catalogue rather than sampled negatives, tunes on a separate validation fold, reports intervals and paired tests, and keeps results obtained under a different protocol out of the comparable table. It also states the limits of its split: chronology is enforced within each reviewer key, not across the dataset (Section~\ref{sec:protocol}). Its distinguishing property is not scale but construction evidence for a market that has none: measured entity-resolution error, a measured lower bound on identifier collisions, and a benchmark that is repeated under a stricter identifier definition.

\section{ViHoRec Dataset}
\label{sec:dataset}

This section describes how the data were collected and hotel listings resolved (Section~\ref{sec:construction}), the quality checks applied before release (Section~\ref{sec:qc}), and the released files and their statistics (Section~\ref{sec:data-description}).

\subsection{Data Collection and Construction}
\label{sec:construction}

\subsubsection{Data collection}
\label{lbl:Collectdata}
\textbf{Platforms and destinations.} Selection was purposive, not a probability sample. We chose three booking platforms with large Vietnamese user bases and public review listings: Booking.com, Traveloka, and Ivivu. We then chose nine leisure destinations with high public review volume: Đà Lạt, Nha Trang, Đà Nẵng, Vũng Tàu, Phan Thiết, Phú Quốc, Hội An, Huế, and Quy Nhơn. The released catalogue holds between 41 (Quy Nhơn) and 98 (Đà Lạt) hotels per destination.

\textbf{Hotels and reviews.} Within each destination, the hotels are those that the platform's own destination listing returned at crawl time; we did not short-list hotels by hand. For each listed hotel, the crawler kept every review that the listing exposed. No cap was applied per hotel or per destination, and reviews were not subsampled. Where a platform offers a review-listing API, we read its JSON responses; otherwise we parse the public HTML. The raw interaction fields are reviewer display name, location, hotel name, rating, and review date. Content metadata, which no platform exposes through an API, was collected semi-manually for hotels that appear in the interaction data.

\textbf{Time frame.} Crawling was carried out in 2023 and ended no earlier than 9 December 2023, the date of the latest collected review. Review dates span 2011-10-15 to 2023-12-09. The span differs by platform: Ivivu reviews start in 2011, Traveloka reviews in 2015, and Booking.com reviews in January 2020, consistent with Booking.com listing only recent reviews at crawl time. The exact calendar start and stop of each crawl job were not logged.

\textbf{Selection bias.} The corpus covers leisure destinations with public review volume, on three platforms, during the period each platform exposes. It makes no coverage claim for all Vietnamese hotels: business-travel cities, budget properties without online reviews, and travellers who do not post reviews are under-represented. Before 2020, every interaction comes from Ivivu or Traveloka, so cross-period comparisons also compare platforms.

\subsubsection{Normalization and hotel entity resolution}
\label{lbl:entity}
A hotel may appear under different name spellings across websites (differing diacritics, presence/absence of prefixes such as ``Khach san''/``Hotel''/``Khu nghi duong'', differing word order, differing punctuation). Naive exact-string matching would treat such variants as distinct hotels and fragment the interaction graph. We assign each listing a city- and type-aware key. Each name is lower-cased and stripped of accents and punctuation; generic words and property-type words are removed, and the remaining tokens are sorted so that word order does not matter. ``Khách sạn Pullman Vũng Tàu'' thus becomes ``pullman tau vung'', the same list as ``Pullman Vung Tau'' and as the reordering ``Vũng Tàu Pullman''. The hyphen in ``Raon Hotel - STAY 24H'' is read as a space, leaving ``24h raon stay'' after ``hotel'' is removed. The city is stored as its own field, and city tokens are dropped from the name unless doing so would leave no tokens: dropping ``vung'' and ``tau'' from the Pullman list leaves ``pullman''. A property type is read from the name when one is stated. An unstated type stays compatible with a stated type, so ``Khách sạn Pullman Vũng Tàu'' (hotel) and ``Pullman Vung Tau'' (no type stated) both resolve to hotel. Two different stated types stay separate: ``Khách sạn Sea Links Beach Phan Thiết'' and ``Sea Links Beach Villas Phan Thiết'' share Phan Thiết and the tokens ``beach links sea'', and they remain distinct because one states hotel and the other villa. Listings that share a city, the same discriminative tokens, and compatible property types are merged into one \texttt{hotel ID}, so the Pullman spellings share the key ``vung tau | pullman | hotel''. The listings produced here are the input to the pair-level check in Section~\ref{lbl:er-pairs}.

\subsection{Quality Control and Validation}
\label{sec:qc}

Data quality is a critical factor for any recommendation dataset, so we perform a \emph{quantitative} quality-control process. We disclose one validity threat up front: because identifiers derive from reviewer display names---a low-cardinality string space in which 357 crawl-time placeholders (1.95\%) were later dropped---distinct individuals sharing an abbreviated name may be merged, so reviewer counts are approximate. We treat \texttt{user ID} as a \emph{pseudonymous reviewer key}, not a person. Table~\ref{tab:qc} summarizes the checks.

\begin{table}[ht]
\centering
\caption{Summary of the ViHoRec quality-control metrics on the released corpus.}
\label{tab:qc}
\small
\setlength{\tabcolsep}{4pt}
\begin{tabularx}{\columnwidth}{@{}>{\raggedright\arraybackslash}X r@{}}
\toprule
\textbf{Check} & \textbf{Result} \\
\midrule
Raw interactions & 18{,}274 \\
Placeholder reviewer names (357 found; 356 dropped) & 1.95\% \\
Other fields missing & 0.0\% \\
Exact duplicates (removed) & 7 (0.038\%) \\
Near-duplicates (reviewer\,+\,hotel\,+\,date) & 11 (0.060\%) \\
Invalid / out-of-range ratings & 0 \\
Unparsable dates & 0 \\
Raw names $\rightarrow$ canonical hotels & 581 $\rightarrow$ 560 \\
\quad name variants merged & 21 (3.61\%) \\
Hotels on $\geq 2$ platforms & 81 \\
Hotels with conflicting location & 0 \\
\midrule
Interactions after cleaning & 17{,}911 \\
Reviewer keys / hotels & 6{,}822 / 560 \\
\bottomrule
\end{tabularx}
\end{table}

\subsubsection{Duplicates and completeness}

We detect duplicates at two levels. An \emph{exact duplicate} repeats the full tuple (reviewer display name, hotel name, rating, review date): 7 extra rows (0.038\%) were removed. A \emph{near-duplicate} repeats the reviewer key, the canonical hotel, and the calendar day, so the rating may still differ. This broader check flags 11 extra rows (0.060\%), and 7 of those 11 are the exact duplicates already removed. The other 4 stay on one platform, keep the same hotel spelling, and carry a second rating. One display name, for example, has two Ivivu rows for ``Khách sạn Fusion Suites Vũng Tàu'' on 2020-05-30, scored 6 and 7. We retain those 4 rows because each pair records two ratings.

Only the reviewer-name field has missing values. In 357 raw rows (1.95\%) the platform showed a placeholder instead of a name (\texttt{Không tên}, ``no name'', \texttt{Guest}, and variants). A placeholder would merge many travellers into one reviewer key, so we drop 356 of these rows before release. Hotel name, location, rating, date, and source have no empty values.

\subsubsection{Reviewer-key reliability}
\label{sec:user-identity}
\begin{figure}[!ht]
    \centering
    \includegraphics[width=\columnwidth]{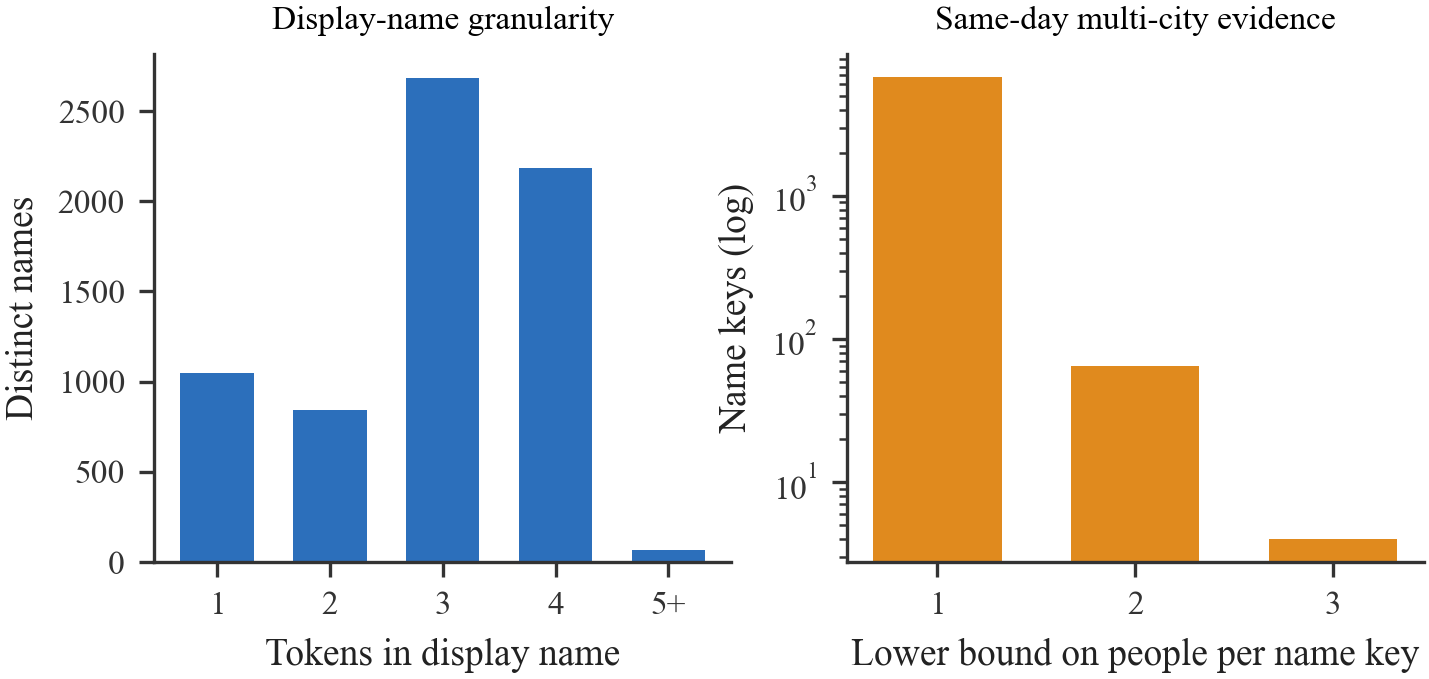}
    \caption{Reviewer-key reliability: name-length mass, collision floor from same-day multi-city events, and sensitivity of counts under stricter identity filters.}
    \label{fig:user-identity}
\end{figure}

HMAC of a display name is a reviewer key, not a recovered platform account. Figure~\ref{fig:user-identity} and a same-day multi-city audit give a \emph{floor} on collisions: 69 name keys (1.01\%) review hotels in two cities on the same calendar day, covering 3{,}513 interactions (19.6\%). Counting one traveller per key plus the extra travellers that these collisions prove gives at least 6{,}895 people behind the 6{,}822 keys; the true number is higher, because collisions without a same-day multi-city event are invisible. Conversely, a traveller who uses different display names on different platforms is split across keys. Single-token names account for 15.3\% of keys but 40.5\% of interactions; the longest remaining history is 210. Per-key history lengths are inflated wherever a common Vietnamese name absorbs several travellers.

To bound this identity limitation and to give a reliability range for the benchmark, we recompute the min-$k{=}4$ pool under two stricter checks. That pool is the set of reviewer keys with at least four interactions, and the released split keeps all 798 of them. Dropping display names of at most three characters, such as ``Anh'' and ``Lan'', leaves 688 keys. Requiring at least two whitespace-separated tokens leaves 493 keys: ``Hoang'' is longer than three characters and is still removed, because it is a single token. Also dropping keys with a proven collision---the same name reviews hotels in two cities on one calendar day---leaves 481 keys. ``Le H. T.'' has several tokens and is removed only by that collision rule.

\subsubsection{Consistency}

We check every rating and store it as a float on the $[1,10]$ scale. A token with an extra character, such as \texttt{8..5}, is repaired to $8.5$. Null and out-of-scale values are caught by the same check and removed. Next, we check the review-date field and parse it to a calendar date. Values that cannot be parsed are removed. The observed span is 2011-10-15 to 2023-12-09.

After entity resolution (Section~\ref{lbl:entity}), we also check location consistency: if one \texttt{hotel ID} is assigned more than one distinct city, that identifier is flagged for review. Three annotators then inspect the flagged listings independently, each working on a single platform (Booking.com, Traveloka, or Ivivu) and without seeing the others' labels. A listing that still maps to two cities after that review is not released as one hotel.

\subsubsection{Cross-platform entity resolution}
\label{lbl:er-pairs}
The same hotel can appear on Booking.com, Traveloka, and Ivivu under different names. To measure how well the resolver reconciles them, we label 426 candidate listing pairs drawn from four strata (Table~\ref{tab:er}). Among pairs that neither matcher joins, a high-similarity pool keeps pairs whose token Jaccard or character-sequence ratio passes a fixed cutoff; the second and third strata are drawn from this pool.

\begin{itemize}
    \item \textbf{Algorithm merges} are the exhaustive stratum built by Algorithm~\ref{alg:algo-merge}: every pair joined by either key is kept, so both matchers are scored on the same set.
    \item \textbf{Highly similar names} are the highest-Jaccard or character-sequence-ratio pairs from the remaining unmerged pool that satisfy either cutoff, and are labelled exhaustively; ``Khu nghỉ dưỡng Hải Dương Intourco Vũng Tàu'' and ``Khu nghỉ dưỡng Intourco Vũng Tàu'' sit at the top of this slice.
    \item \textbf{Hard negatives} are the remaining pairs from the high-similarity pool; ``Khách sạn Riva Vũng Tàu'' and ``Khách sạn Romeliess Vũng Tàu'' share a city and a similar name pattern but rank below the highly similar slice.
    \item \textbf{Random negatives} are drawn uniformly from all remaining listing pairs and estimate how often a duplicate escapes every similarity cutoff.
\end{itemize}

\begin{algorithm}[!ht]
\footnotesize
\caption{Algorithm-merge stratum. A pair enters the stratum when either key joins the two listings, and every such pair is labelled.}
\label{alg:algo-merge}
\begin{algorithmic}[1]
\Require listings $L$; name-only key $k_{\mathrm{name}}$; city- and type-aware key $k_{\mathrm{city}}$
\State $\mathcal{M} \gets \varnothing$
\For{each unordered pair $(a,b)$ of distinct listings in $L$}
  \If{$k_{\mathrm{name}}(a)=k_{\mathrm{name}}(b) \lor k_{\mathrm{city}}(a)=k_{\mathrm{city}}(b)$}
    \State $\mathcal{M} \gets \mathcal{M} \cup \{(a,b)\}$
  \EndIf
\EndFor
\State Label every pair in $\mathcal{M}$ \Comment{exhaustive; no subsampling}
\State \Return $\mathcal{M}$
\end{algorithmic}
\end{algorithm}

\textbf{Labelling.} Two humans label all 426 pairs independently, blind to matcher decisions and stratum, with the rules in Appendix~\ref{app:llm-prompt}. Annotator~A labelled the pairs in two blind sessions of 130 and 296 pairs; annotator~B, who did not build the resolver, labelled all 426. Disagreements go to a third annotator~C, who also did not build the resolver, and the adjudicated human labels are the gold standard.

\textbf{Agreement.} The two humans agree on 424 of the 426 pairs (99.5\%), with Cohen's $\kappa{=}0.95$ (95\% bootstrap CI $[0.87, 1.00]$). Because only 22 of the 426 pairs are duplicates, $\kappa$ is sensitive to the prevalence of negatives, so we also report the prevalence-adjusted PABAK (0.99), positive agreement (0.95), and negative agreement (1.00). Both disagreements lie in the highly similar stratum, and in both annotator~A judged the pair to be one hotel and annotator~B two: ``Khu nghỉ dưỡng Hải Dương Intourco Vũng Tàu'' / ``Khu nghỉ dưỡng Intourco Vũng Tàu'' and ``Khách sạn TTC - Ngọc Lan Đà Lạt'' / ``Khách sạn TTC – Đà Lạt''. The hard-negative, random-negative, and matcher-merge strata contain no disagreement. Annotator~C, consulting public listings, ruled the Intourco pair two different resorts (different addresses on the same street, different star ratings, and separate platform listings) and the TTC pair one hotel listed under two name variants.

\begin{table}[t]
\centering
\caption{Entity-resolution results on 426 candidate pairs with adjudicated two-annotator gold labels. Baseline keys listings on the normalised name alone; Ours adds city and property type (Section~\ref{lbl:entity}). Lo/Hi: exact Clopper--Pearson 95\% interval. Precision covers every proposed merge; recall is computed over the 22 gold duplicates in the sample, not the full pair population. Bold: the better matcher on each error measure.}
\label{tab:er}
\footnotesize
\setlength{\tabcolsep}{3pt}
\begin{tabular*}{\columnwidth}{@{\extracolsep{\fill}}lccc ccc@{}}
\toprule
 & \multicolumn{3}{c}{\textbf{Baseline}} & \multicolumn{3}{c}{\textbf{Ours}} \\
 & \multicolumn{3}{c}{(name only)} & \multicolumn{3}{c}{(city + type)} \\
\cmidrule(lr){2-4}\cmidrule(lr){5-7}
 & Est. & Lo & Hi & Est. & Lo & Hi \\
\midrule
Merges proposed & 22 & & & 21 & & \\
False merges & 5 & & & \textbf{0} & & \\
Precision & 0.77 & 0.55 & 0.92 & \textbf{1.00} & 0.84 & 1.00\\
False-merge rate & 0.23 & 0.08 & 0.45 & \textbf{0.00} & 0.00 & 0.16 \\
\midrule
True duplicates in sheet & 22 & & & 22 & & \\
False splits (missed) & 5 & & & \textbf{1} & & \\
Sample recall & 0.77 & 0.55 & 0.92 & \textbf{0.95} & 0.77 & 1.00\\
False-split rate & 0.23 & 0.08 & 0.45 & \textbf{0.05} & 0.00 & 0.23 \\
\bottomrule
\end{tabular*}
\end{table}

\textbf{Error rates.} Table~\ref{tab:er} reports both matchers on the same gold labels. The name-only baseline makes 5 false merges and misses 5 of the 22 duplicates. Our released resolver proposes 21 merges and none is false; the exact 95\% upper bound on its false-merge rate is 0.16. It misses one of the 22 duplicates, in the highly similar stratum: ``Khách sạn TTC - Ngọc Lan Đà Lạt'' / ``Khách sạn TTC – Đà Lạt''. The hard- and random-negative strata contain no duplicate.

\subsubsection{Manual validation and inter-annotator agreement}
\label{lbl:manual}
To estimate corpus quality beyond the automated checks above, we draw a stratified random sample of 248 records: 186 interactions balanced across Booking.com, Traveloka, and Ivivu, and 62 hotel catalogue entries. Each record receives three binary checklist scores, with $1$ for pass and $0$ for fail. C1 scores field validity (a rating in $[1,10]$, a parseable date, and non-empty identifiers). C2 scores release consistency (the same tuple appears in the released tables with the same rating, date, and source). C3 scores identity plausibility (the reviewer name is not a placeholder and is longer than three characters, and the hotel location is not a multi-city conflict). These three scores are deterministic pass rates. Three annotators then label each of the 62 hotels independently as a valid catalogue entry or not: the name and city are non-empty, the hotel identifier is unique, and the hotel has at least one interaction. The decision rules are in Appendix~\ref{app:guidelines}.

On this sample, C1 and C2 are positive for every record. C3 is positive for 89.5\% of records (222 of 248). The 26 failures are all interactions: 26 of 186 interaction rows fail the identity check, and none of the 62 hotels do. On the hotel panel, the three annotators agree before adjudication on 61 of 62 entries, a percent agreement of 98.4\%. The remaining entry is valid after adjudication, and all 62 released hotel labels are valid. We therefore report percent agreement. C1--C3 stay pass rates, because each score is a checklist criterion rather than an annotator.

\subsection{Dataset Description}
\label{sec:data-description}

The ViHoRec dataset is released as CSV files encoded in UTF-8 and consists of two complementary parts: (i) \emph{interaction data} for collaborative filtering and (ii) \emph{hotel content metadata} for content-based recommendation. Table~\ref{tab:files} summarizes the files and their schemas; Appendix~\ref{app:schema} provides full field definitions and sample records.

\begin{table}[!ht]
\centering
\caption{Files in the ViHoRec dataset and their schemas.}
\label{tab:files}
\small
\setlength{\tabcolsep}{3pt}
\begin{tabularx}{\columnwidth}{@{}l r >{\raggedright\arraybackslash}X@{}}
\toprule
\textbf{File} & \textbf{Rows} & \textbf{Fields} \\
\midrule
\texttt{interactions} & 17{,}911 & \texttt{user ID}, \texttt{hotel ID}, \texttt{rating}, \texttt{date}, \texttt{source} \\
\texttt{users} & 6{,}822 & \texttt{user ID}, \texttt{N-interactions} \\
\texttt{hotels} & 560 & \texttt{hotel ID}, \texttt{name}, \texttt{location} \\
content metadata & 309 & 11 attributes (see text) \\
\bottomrule
\end{tabularx}
\end{table}

\textbf{Interaction data.} Each row in \texttt{interactions} is one review by a reviewer key for a hotel: a salted-HMAC \emph{pseudonymous} identifier (\texttt{user ID}), a canonical hotel identifier (\texttt{hotel ID}), a \texttt{rating} score on the $[1,10]$ scale (mean $7.58$), a \texttt{date} (roughly 2011-10-15 to 2023-12-09), and a \texttt{source} indicating the origin platform. The three platforms contribute, respectively: Booking.com 7{,}597; Traveloka 6{,}273; Ivivu 4{,}404 raw reviews. We treat \texttt{user ID} as a reviewer-name key, not a person identity (Section~\ref{sec:qc}).

\textbf{Content metadata.} For 309 hotels we collect 11 attributes: name, location, overall rating, number of reviews, price, facilities, quality, distance to center, surrounding places (Around), vicinity, and booking link. The four textual attribute groups (Facilities, Around, Vicinity, Price) are used as features for content-based recommendation.

\subsubsection{Dataset statistics}
\label{lbl:Datasetstatistics}

\begin{figure}[!ht]
    \centering
    \begin{subfigure}[t]{0.48\columnwidth}
        \includegraphics[width=\linewidth]{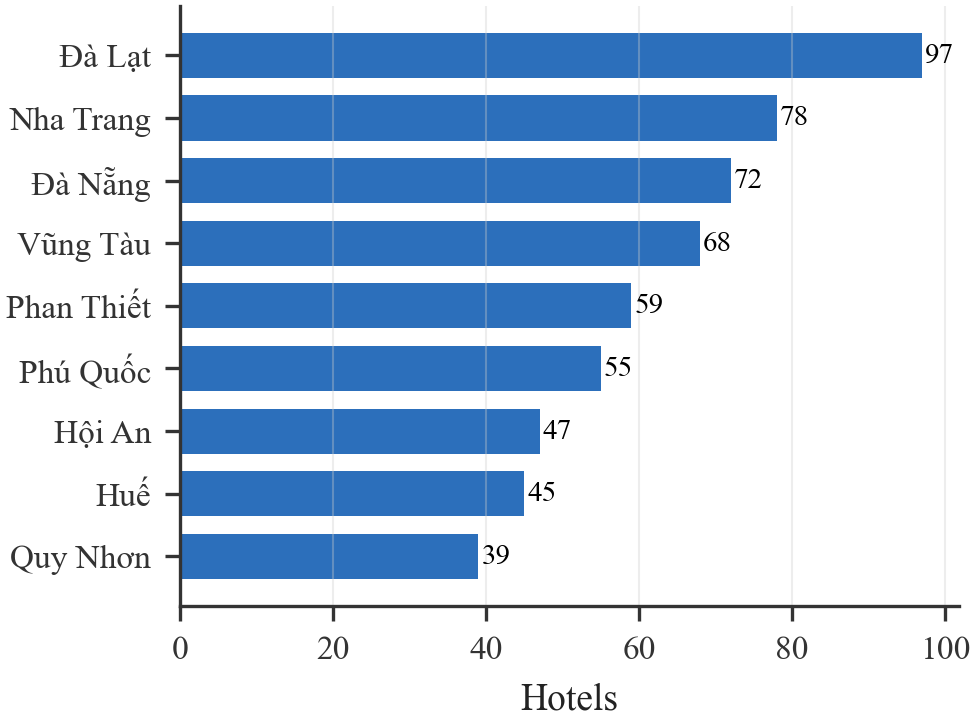}%
        \caption{Number of hotels per location.}
        \label{fig:HotelByLoc}
    \end{subfigure}%
    \hfill
    \begin{subfigure}[t]{0.48\columnwidth}
        \includegraphics[width=\linewidth]{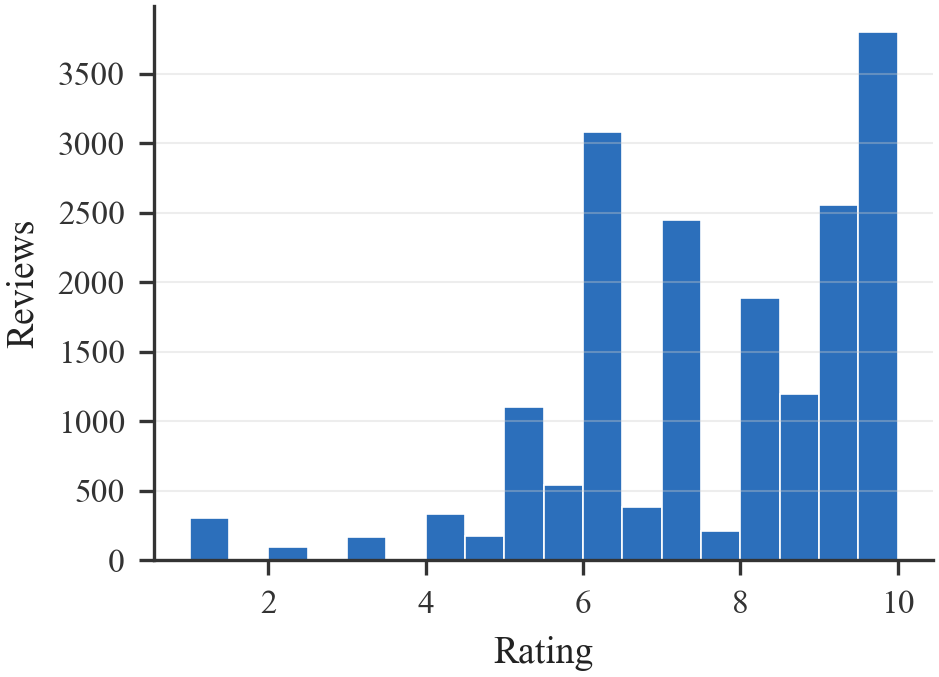}%
        \caption{Distribution of rating values.}
        \label{fig:RatingDist}
    \end{subfigure}
    \caption{Distribution of hotels across locations and distribution of rating scores.}
    \label{fig:CombinedRating}
\end{figure}

Figure~\ref{fig:CombinedRating}a shows that the catalogue covers nine destinations, and that no city holds a majority of hotels. Đà Lạt is the largest group, with 98 hotels (17.5\%). Figure~\ref{fig:CombinedRating}b shows that ratings on the $[1,10]$ scale concentrate at $6$ and above. The mean is $7.58$ (std $1.95$), the median is $8.0$, and the mode is $10$ (2{,}729 reviews, 15.2\%); scores of at least $6$ account for 15{,}307 of 17{,}911 reviews (85.5\%).

\begin{figure}[!ht]
    \centering
    \begin{subfigure}[t]{0.48\columnwidth}
        \includegraphics[width=\linewidth]{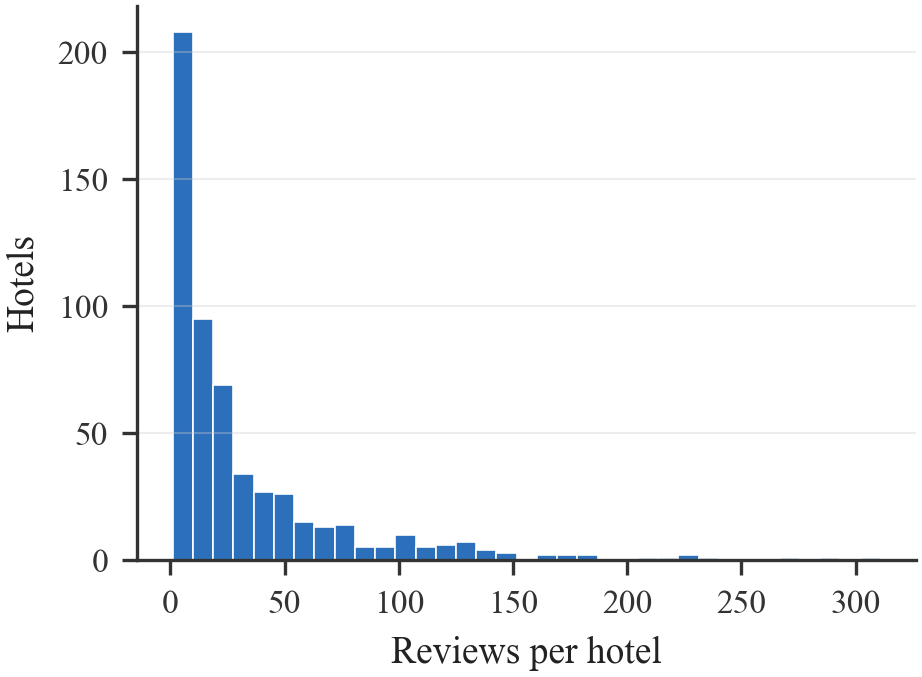}%
        \caption{Number of reviews per hotel.}
        \label{fig:HotelRating}
    \end{subfigure}%
    \hfill
    \begin{subfigure}[t]{0.48\columnwidth}
        \includegraphics[width=\linewidth]{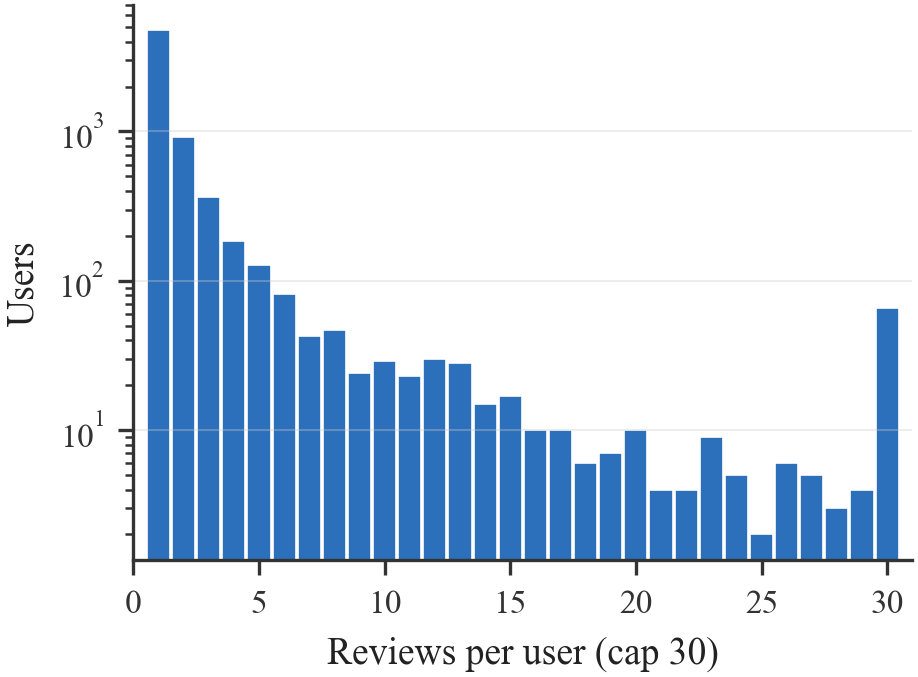}%
        \caption{Number of reviews per reviewer key.}
        \label{fig:CusRating}
    \end{subfigure}
    \caption{Review counts in the full corpus. (a)~560 hotels. (b)~6{,}822 reviewer keys (log scale; histories of 30 or more pooled in the last bar).}
    \label{fig:ReviewCounts}
\end{figure}

Figure~\ref{fig:HotelRating} shows that review counts per hotel are right-skewed. Across all 560 hotels the median is 17 reviews and the mean is 32.0; 123 hotels (22.0\%) have at most five reviews, whereas 43 hotels (7.7\%) have at least 100 and the busiest hotel has 308. Figure~\ref{fig:CusRating} shows that reviewer-key histories are shorter still: of 6{,}822 keys, 4{,}741 (69.5\%) have exactly one review, and the last bar pools 65 keys (1.0\%) with 30 or more; the longest history in the corpus is 210 reviews.

\begin{figure}[!ht]
    \centering
    \includegraphics[width=\columnwidth]{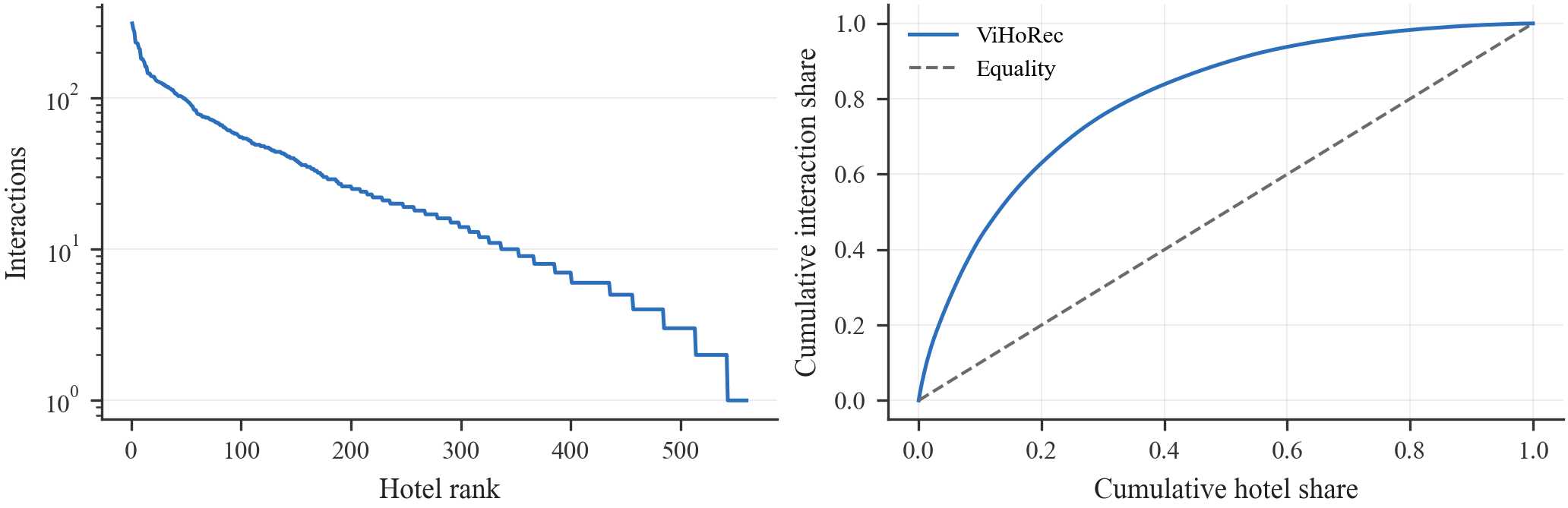}
    \caption{Item-popularity long tail (left, log scale) and Lorenz curve (right). Popularity is heavily concentrated in a few hotels.}
    \label{fig:longtail}
\end{figure}

Table~\ref{tab:charac} and Figure~\ref{fig:longtail} summarize that corpus as an interaction matrix. The full corpus is 99.53\% sparse. Item popularity is concentrated: the Gini coefficient is 0.593, and the top 20\% of hotels account for 62.3\% of interactions. The mean history length is 2.63 interactions per reviewer key (median 1).

\begin{table}[!ht]
\centering
\caption{Characterization of the full ViHoRec corpus after placeholder drop.}
\label{tab:charac}
\small
\setlength{\tabcolsep}{4pt}
\begin{tabularx}{\columnwidth}{@{}>{\raggedright\arraybackslash}X r@{}}
\toprule
\textbf{Property} & \textbf{Value} \\
\midrule
Reviewer keys / hotels / interactions & 6{,}822 / 560 / 17{,}911 \\
Sparsity & 99.53\% \\
Interactions per reviewer key (mean / median) & 2.63 / 1 \\
Longest reviewer-key history & 210 \\
Interactions per hotel (mean) & 32.0 \\
Item-popularity Gini & 0.593 \\
Top-20\% hotels' interaction share & 62.3\% \\
Single-interaction reviewer keys & 69.5\% \\
Rating (mean $\pm$ std) & 7.58 $\pm$ 1.95 \\
\bottomrule
\end{tabularx}
\end{table}

\section{Adaptive Hybrid: A Diagnostic Baseline}
\label{sec:methods}
The full ViHoRec corpus is cold-start dominated as a \emph{dataset property}: 69.5\% of reviewer keys have a single interaction and are excluded from the public split. On the split itself, collaborative filters still have short training histories (minimum two), and only 53.5\% of benchmark hotels have text metadata. Adaptive Hybrid is a closed-form probe of one question about the release: does the metadata add anything to a collaborative ranker, and at which history lengths? It is a diagnostic baseline, not a proposed method. It late-fuses UserKNN and Content-TFIDF with a weight that depends on history length.

\begin{figure}[t]
    \centering
    \includegraphics[width=0.75\columnwidth]{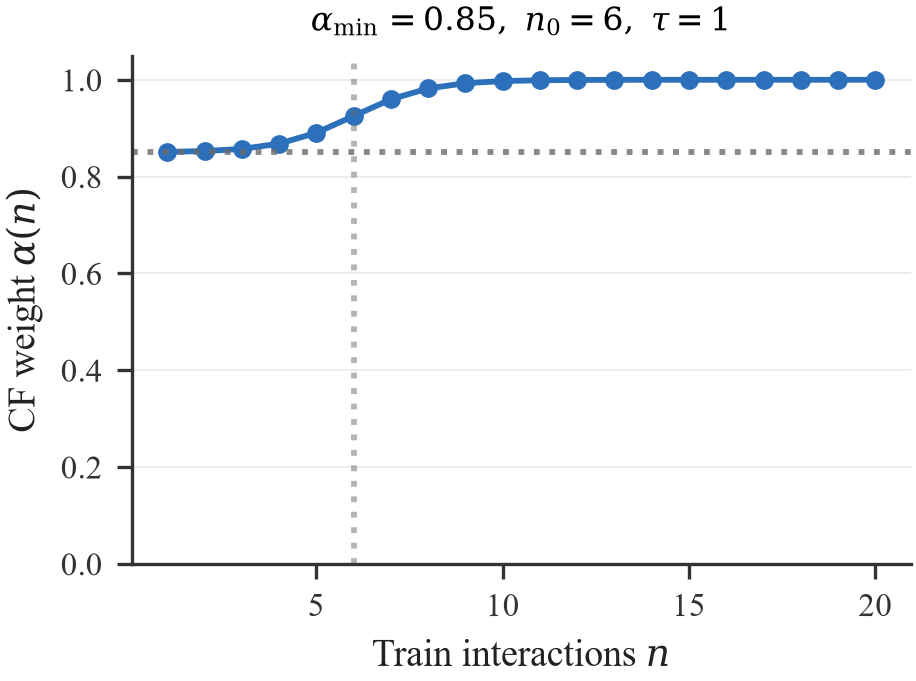}
    \caption{Bounded CF weight $\alpha(n)$ used by Adaptive Hybrid after validation selection ($\alpha_{\min}{=}0.85$, $n_0{=}6$, $\tau{=}1$). Short histories keep a content residual; longer histories approach pure CF.}
    \label{fig:adaptive-alpha}
\end{figure}

Let $n_u$ be the number of training interactions for reviewer key $u$, and let $\tilde{s}_{\mathrm{CF}}$ and $\tilde{s}_{\mathrm{CB}}$ denote UserKNN and Content-TFIDF scores after per-key min--max normalisation over items. Adaptive Hybrid ranks by
\begin{equation}
s_{\mathrm{AH}}(u,i)=\alpha(n_u)\,\tilde{s}_{\mathrm{CF}}(u,i)+\bigl(1-\alpha(n_u)\bigr)\,\tilde{s}_{\mathrm{CB}}(u,i),
\label{eq:adaptive-hybrid}
\end{equation}
with a bounded logistic gate
\begin{equation}
\alpha(n)=\alpha_{\min}+(1-\alpha_{\min})\,\sigma\!\left(\frac{n-n_0}{\tau}\right),
\label{eq:alpha-gate}
\end{equation}
where $\sigma(x)=(1+e^{-x})^{-1}$. Short histories receive a larger content residual ($\alpha$ closer to $\alpha_{\min}$); longer histories approach pure CF ($\alpha\to 1$). We select $(\alpha_{\min}, n_0, \tau)$ on validation only; the chosen setting is $\alpha_{\min}{=}0.85$, $n_0{=}6$, $\tau{=}1$ (Figure~\ref{fig:adaptive-alpha}). The CF floor is essential: an unconstrained gate with $\alpha_{\min}{=}0$ (content-dominant for short histories) collapses shortest-bucket Recall@10 to $0.044$, because Content-TFIDF remains weaker than UserKNN even at history length~2. Items without metadata fall back to CF scores so that incomplete coverage does not zero out ranking.

The gate is closed-form and dependency-free: it uses only the released interaction counts and hotel metadata. The selected $\alpha_{\min}{=}0.85$ keeps at least 85\% of the weight on UserKNN for every key, so the content signal can only reorder items that UserKNN already scores closely. Section~\ref{sec:results} tests it against UserKNN on the full test fold, and Section~\ref{sec:coldstart} tests it within each history bucket.

\section{Experiments}
\label{sec:experiments}

This section fixes the evaluation setup (Section~\ref{sec:benchmark}), reports the comparable results (Section~\ref{sec:results}), and tests their robustness (Section~\ref{sec:analysis}).

\subsection{Experimental Setup}
\label{sec:benchmark}

Three questions organize the experiments, on the corpus characterized in Section~\ref{lbl:Datasetstatistics}. Under one protocol, which model families rank the held-out hotel best when training histories are short? Do those conclusions survive a stricter reviewer-key definition and the correction of known entity-resolution errors? Does history-gated fusion with the released metadata differ from UserKNN at any history length? This subsection fixes the protocol. Section~\ref{sec:results} answers the first question, and Section~\ref{sec:analysis} answers the other two.

\subsubsection{Metrics and Protocol}
\label{sec:setup-protocol}
\label{sec:protocol}

\textbf{Split.} For each reviewer key with at least four interactions, the last interaction is the test item, the second-last the validation item, and the rest (at least two) the training history. This temporal leave-one-out split, the \emph{freeze}, has 798 keys, 535 hotels, and 8{,}645/798/798 training/validation/test interactions (97.60\% sparsity); single-interaction keys (69.5\% of the corpus) are released but not scored.

\textbf{Scope of the temporal ordering.} Chronology is enforced within each key, with no global cutoff, so one key's training data may postdate another key's test event. The task is therefore per-key next-item ranking, not a prospective forecast~\cite{meng2020splitting,ji2023leakage}.

\textbf{Candidates, repeated visits, and ties.} Each key ranks the full catalogue minus its training hotels, with no sampled negatives~\cite{krichene2020sampled}; the validation hotel remains a candidate. After exact-duplicate removal (Section~\ref{sec:qc}), repeated visits are kept as separate rows (290 keys). For 74 keys (9.3\%) the test hotel already appears in training, so masking caps attainable Recall@10 at $724/798{=}0.907$ for every model; for another 22 keys it never occurs in training. Same-day interactions are ordered by hotel identifier, which decides the holdout for 9 keys.

\begin{table}[ht]
\centering
\caption{Evaluation protocol shared by all comparable results. Numbers obtained under other settings appear only in Table~\ref{tab:reference}.}
\label{tab:protocol}
\small
\setlength{\tabcolsep}{4pt}
\begin{tabularx}{\columnwidth}{@{}>{\raggedright\arraybackslash}p{2.35cm} >{\raggedright\arraybackslash}X@{}}
\toprule
\textbf{Element} & \textbf{Setting} \\
\midrule
Training data & \texttt{train.csv} only\\
Validation & \texttt{val.csv}; model selection only, never training labels\\
Test-time history & training interactions only; validation item not appended\\
Candidates & 535 hotels minus the key's training hotels; full ranking\\
Test fold & scored once, after all selection is fixed\\
Statistics & per-key bootstrap 95\% CI; paired Wilcoxon vs.\ UserKNN \\
\bottomrule
\end{tabularx}
\end{table}

\textbf{Selection and statistics.} Validation selects the Hybrid-fixed weight $\lambda$, the Adaptive Hybrid parameters $(\alpha_{\min}, n_0, \tau)$, the EASE and SANSA regularizers, and the early-stopping epoch or checkpoint of each neural model; the test fold is scored once. We report percentile-bootstrap 95\% confidence intervals over keys (10{,}000 resamples) and two-sided paired Wilcoxon signed-rank tests on per-key Recall@10 against UserKNN, with Holm correction for the bucket-level tests of Section~\ref{sec:coldstart}.

\textbf{Metrics.} The task is implicit next-item ranking: the holdout hotel is the only relevant item, and every rating counts as a positive, because the label is the visit. We compute MAP, NDCG, Precision, and Recall at $K \in \{5, 10\}$ and MRR over the full ranking~\cite{chen2017performance}, and report MRR, NDCG@10 (N@10), and Recall@10 (R@10). A companion split that counts only holdouts rated at least 7 as relevant is released with the same training fold.

\subsubsection{Baseline models}
\label{sec:baselines}

The closed-form set covers popularity, neighborhood, latent-factor, content, and hybrid ranking. Random is the sanity floor. MostPop ranks hotels by training-set frequency~\cite{cremonesi2010performance}. ItemKNN and UserKNN use cosine similarity~\cite{sarwar2001item,sgardelis2025evaluation}. BPR-MF learns latent factors with Bayesian personalized ranking~\cite{rendle2012bpr,koren2009matrix}. Content-TFIDF concatenates a hotel's Facilities, Around, Vicinity, and Price fields into one lower-cased, tokenized document, represents it as a TF-IDF vector~\cite{manning2008introduction}, and scores it by cosine similarity to the mean of the L2-normalized vectors of the key's training hotels. Text metadata covers 286 of the 535 hotels (53.5\%); Content-TFIDF ranks hotels without text last, and the two hybrids use the collaborative score in place of their missing content score. Hybrid-fixed fuses UserKNN and Content-TFIDF with a global collaborative weight, $\lambda{=}0.7$ on validation. Adaptive Hybrid replaces that constant with the history gate in Section~\ref{sec:methods}. EASE is the closed-form item--item autoencoder~\cite{steck2019ease}, with validation-selected $\lambda{=}1000$.

Neural models are trained on the same freeze. LightGCN~\cite{he2020lightgcn}, NeuMF, and NCF-NeuMF~\cite{he2017ncf} are neural collaborative filters, and SANSA~\cite{spisak2023sansa} is a sparse approximate inverse of the EASE item--item model. GRU4Rec~\cite{hidasi2016gru4rec}, SASRec~\cite{kang2018sasrec}, and SSE-PT~\cite{wu2020ssept} are sequential models. TIGER~\cite{rajput2023tiger}, LETTER~\cite{wang2024letter}, and RPG~\cite{hou2025rpg} are generative retrieval models that decode semantic item identifiers. Implementations are in-repo PyTorch for LightGCN, NeuMF, and GRU4Rec; Microsoft Recommenders~\cite{argyriou2020microsoft} for NCF-NeuMF, SASRec, and SSE-PT; and Cornac~\cite{salah2020cornac} for SANSA, TIGER, LETTER, and RPG. The generative models keep the training configuration of their Cornac implementations; for test scoring we condition on the key's training history only and compute the exact likelihood of every catalogue item instead of the library's default scoring, so that they rank the same masked candidate set as every other model. 

\begin{table*}[ht]
\centering
\caption{Closed-form baselines on the public freeze under the protocol of Table~\ref{tab:protocol}. Shaded: the reference row; bold: best point estimate per column (no method is significantly above UserKNN). $p$: two-sided paired Wilcoxon test on per-key Recall@10; $^{***}\,p{<}0.001$.}
\label{tab:baseline}
\small
\setlength{\tabcolsep}{5pt}
\begin{tabular}{@{}lcccc@{}}
\toprule
\textbf{Method} & \textbf{MRR}$\uparrow$ & \textbf{N@10}$\uparrow$ & \textbf{R@10}$\uparrow$ & \textbf{$p$ vs.\ UserKNN} \\
\midrule
Random          & 0.0123 & 0.0080 & 0.0159 & $<0.001^{***}$ \\
MostPop~\cite{cremonesi2010performance}         & 0.0496 & 0.0539 & 0.1103 & $0.059$ \\
ItemKNN-cosine~\cite{sarwar2001item}  & 0.0426 & 0.0420 & 0.0865 & $<0.001^{***}$ \\
\rowcolor{vihbg}
UserKNN-cosine~\cite{sgardelis2025evaluation}  & 0.0606 & 0.0644 & 0.1291 & --- \\
BPR-MF~\cite{rendle2012bpr}          & 0.0493 & 0.0485 & 0.0977 & $<0.001^{***}$ \\
Content-TFIDF~\cite{manning2008introduction}   & 0.0287 & 0.0252 & 0.0551 & $<0.001^{***}$ \\
Hybrid-fixed ($\lambda{=}0.7$) & 0.0608 & 0.0665 & \textbf{0.1353} & $0.50$ \\
Adaptive Hybrid & \textbf{0.0639} & \textbf{0.0675} & 0.1328 & $0.44$ \\
EASE~\cite{steck2019ease} & 0.0573 & 0.0595 & 0.1203 & $0.21$ \\
\bottomrule
\end{tabular}
\end{table*}

\subsection{Main Results}
\label{sec:results}

All numbers in this subsection follow the single protocol of Table~\ref{tab:protocol} on the public freeze (798 reviewer keys, 535 hotels).

\textbf{Closed-form models.} UserKNN is the strongest pure collaborative filter in the closed-form set (Table~\ref{tab:baseline}): Recall@10 is 0.129, with a 95\% bootstrap interval of $[0.107, 0.153]$. MostPop reaches 0.110 ($p{=}0.059$). ItemKNN (0.087), BPR-MF (0.098), and Content-TFIDF (0.055) lie below UserKNN ($p{<}0.001$). Hybrid-fixed, at the validation-chosen $\lambda{=}0.7$, reaches 0.135 ($p{=}0.50$ vs.\ UserKNN), and the diagnostic Adaptive Hybrid reaches 0.133 ($p{=}0.44$ vs.\ UserKNN; $p{=}0.77$ vs.\ Hybrid-fixed). EASE at $\lambda{=}1000$ reaches 0.120 ($p{=}0.21$). No closed-form method is significantly better than UserKNN.

\textbf{Neural and sequential models.} Table~\ref{tab:CFresult} lists every neural model under the same protocol. Sequential models have the highest Recall@10 on the freeze. SASRec and SSE-PT reach 0.172 at seed 42 ($p{<}0.001$ vs.\ UserKNN); over seeds 42--46 their means are 0.168 $\pm$ 0.006 and 0.168 $\pm$ 0.010 (Appendix~\ref{app:seeds}). SASRec is significantly above UserKNN at every seed (largest $p{=}0.008$); SSE-PT is significant at four of five seeds, and at seed 45 it reaches 0.152 with $p{=}0.055$. The single-seed tie at 0.172 is therefore not a stable property of the two models. GRU4Rec reaches 0.160 ($p{=}0.007$). LightGCN (0.145, $p{=}0.15$), NCF-NeuMF (0.144, $p{=}0.14$), NeuMF (0.123, $p{=}0.54$), and SANSA (0.119, $p{=}0.14$) are not significantly different from UserKNN. Under the same protocol, the generative models TIGER, LETTER, and RPG reach 0.148 ($p{=}0.10$), 0.144 ($p{=}0.26$), and 0.114 ($p{=}0.22$). The SASRec advantage is specific to the freeze: on the strict reviewer-key pool it is not significant (Section~\ref{sec:strict}).

\begin{table*}[!t]
\centering
\caption{Neural models on the public freeze under the protocol of Table~\ref{tab:protocol}. SASRec and SSE-PT are means over seeds 42--46 (per-seed results in Appendix~\ref{app:seeds}), and their $p$ is the largest over those runs; every other learned model is a single run at seed 42, and SANSA is deterministic. Shaded: highest Recall@10; bold: best value per column. $p$: two-sided paired Wilcoxon on per-key Recall@10; $^{**}\,p{<}0.01$. The generative models are scored with exact full-catalogue likelihoods.}
\label{tab:CFresult}
\small
\setlength{\tabcolsep}{5pt}
\begin{tabular}{@{}llcccc@{}}
\toprule
\textbf{Family} & \textbf{Method} & \textbf{MRR}$\uparrow$ & \textbf{N@10}$\uparrow$ & \textbf{R@10}$\uparrow$ & \textbf{$p$ vs.\ UserKNN} \\
\midrule
Reference & UserKNN-cosine & 0.0606 & 0.0644 & 0.1291 & --- \\
\midrule
Graph / CF & LightGCN~\cite{he2020lightgcn} & 0.0635 & 0.0714 & 0.1454 & $0.15$ \\
 & NeuMF~\cite{he2017ncf} & 0.0538 & 0.0604 & 0.1228 & $0.54$ \\
 & NCF-NeuMF~\cite{he2017ncf} & 0.0681 & 0.0726 & 0.1441 & $0.14$ \\
 & SANSA~\cite{spisak2023sansa} & 0.0583 & 0.0598 & 0.1190 & $0.14$ \\
\midrule
Sequential & GRU4Rec~\cite{hidasi2016gru4rec} & 0.0684 & 0.0764 & 0.1604 & $0.007^{**}$ \\
\rowcolor{vihbg}
 & SASRec~\cite{kang2018sasrec} & \textbf{0.0754} & \textbf{0.0839} & \textbf{0.1684} & $0.008^{**}$ \\
 & SSE-PT~\cite{wu2020ssept} & 0.0727 & 0.0818 & 0.1677 & $0.055$ \\
\midrule
Generative & TIGER~\cite{rajput2023tiger} & 0.0623 & 0.0688 & 0.1479 & $0.096$ \\
 & LETTER~\cite{wang2024letter} & 0.0635 & 0.0702 & 0.1441 & $0.26$ \\
 & RPG~\cite{hou2025rpg} & 0.0585 & 0.0597 & 0.1140 & $0.22$ \\
\bottomrule
\end{tabular}
\end{table*}

\textbf{Reference numbers under the library-default evaluation.} The Cornac leave-last-out evaluator, the library default, scored TIGER, LETTER, and RPG before the protocol of Table~\ref{tab:protocol} was fixed. That setting differs from Table~\ref{tab:protocol} in three ways. The validation interaction is appended to the test-time history, so the model sees one more interaction than every other row. The evaluator also ranks every known hotel without masking the key's training hotels, and TIGER and LETTER score only the hotels reached by a beam of 20, so the 74 repeat-visit targets of Section~\ref{sec:protocol} remain reachable. Finally, the evaluator drops hotels that never occur in training from the test sessions; for the 22 keys whose test hotel is such a hotel, the validation hotel silently became the target. Table~\ref{tab:reference} keeps those numbers for traceability and decomposes the change. Rerunning the earlier setting reproduces TIGER (0.158 vs.\ 0.157 reported) and RPG (0.145) closely. LETTER's rerun gives 0.159 against the reported 0.147, inside the rerun's 95\% interval $[0.134, 0.184]$. Code, seed, and configuration are unchanged, so we treat the gap as run-to-run variation of single-seed training, and only the rerun enters rows 3 and 4. Moving from the rerun to the single protocol lowers Recall@10 by 0.010 (TIGER), 0.015 (LETTER), and 0.031 (RPG). Most of the drop comes from exact scoring with the mask (row 3); removing the validation interaction from the history then changes Recall@10 by at most 0.005, in either direction. The setting also changes the conclusion: in the rerun of the earlier setting TIGER and LETTER would appear above UserKNN ($p{=}0.02$ and $0.04$), but under the protocol no generative model differs significantly from UserKNN. The earlier numbers are not comparable with Table~\ref{tab:CFresult}, carry no significance test, and are not used in any ranking claim.

\begin{table}[t]
\centering
\caption{Recall@10 of the generative models as the test-time setting moves to the protocol of Table~\ref{tab:protocol}. Rows 1--2: Cornac's default leave-last-out evaluation. Row 3: exact full-catalogue scoring with masking, validation item still in the history. Row 4: the protocol; only this row is comparable. Trained models are identical across rows.}
\label{tab:reference}
\footnotesize
\setlength{\tabcolsep}{3pt}
\begin{tabularx}{\columnwidth}{@{}>{\raggedright\arraybackslash}X ccc@{}}
\toprule
\textbf{Test-time setting} & \textbf{TIGER} & \textbf{LETTER} & \textbf{RPG} \\
\midrule
Library default (reported) & 0.157 & 0.147 & 0.145\\
Library default (rerun) & 0.158 & 0.159 & 0.145\\
+ exact scoring, mask & 0.143 & 0.142 & 0.115 \\
\rowcolor{vihbg}
+ train-only history & 0.148 & 0.144 & 0.114 \\
\bottomrule
\end{tabularx}
\end{table}

\subsection{Results Analysis}
\label{sec:analysis}

Section~\ref{sec:results} ranks model families on the public freeze. This subsection asks whether that ranking survives a stricter reviewer-key definition and the correction of known entity-resolution errors, and whether the history-bucket differences are distinguishable from noise.

\subsubsection{Sensitivity to the reviewer-key definition}
\label{sec:strict}

Reviewer keys are display names, so the freeze can merge travellers who share a name. We rebuild the benchmark on the strict pool of Section~\ref{sec:user-identity}: keys whose display name has at least two tokens and that show no same-day multi-city event, followed by the same min-$k{=}4$ split, masking, and metrics as the freeze. The filter removes whole keys, so each of the 481 strict keys is also a freeze key with the same history and the same test hotel; 183 of the 798 freeze keys and 156 of the 481 strict keys have a training length of two. Table~\ref{tab:strict} reports both pools. The strict catalogue has 396 hotels, so the models retrained on the strict pool rank a shorter list, and a random ranker's expected Recall@10 rises from about $10/535 \approx 0.019$ to about $10/396 \approx 0.025$; comparisons within a pool are unaffected, but the two pools are not on one scale. The bootstrap 95\% interval of UserKNN is $[0.107, 0.153]$ on the freeze and $[0.091, 0.148]$ on the strict pool.

\begin{table}[ht]
\centering
\caption{Public freeze versus the strict reviewer-key pool (Section~\ref{sec:user-identity}); models are retrained on each pool. SASRec and SSE-PT are five-seed means (Appendix~\ref{app:seeds}). Chance: expected Recall@10 of a random ranker. Shaded: reference; bold: best per pool. The catalogues differ, so compare within, not across, pools.}
\label{tab:strict}
\small
\setlength{\tabcolsep}{3pt}
\begin{tabular}{@{}lcccc@{}}
\toprule
 & \multicolumn{2}{c}{\textbf{Public freeze}} & \multicolumn{2}{c}{\textbf{Strict pool}} \\
\cmidrule(lr){2-3}\cmidrule(lr){4-5}
 & \textbf{R@10} & \textbf{N@10} & \textbf{R@10} & \textbf{N@10} \\
\midrule
Keys / hotels & \multicolumn{2}{c}{798 / 535} & \multicolumn{2}{c}{481 / 396} \\
Interactions & \multicolumn{2}{c}{10{,}241} & \multicolumn{2}{c}{3{,}436} \\
History mean / med. & \multicolumn{2}{c}{12.8 / 7} & \multicolumn{2}{c}{7.1 / 5} \\
Chance (10 / cat.) & 0.019 & --- & 0.025 & --- \\
\midrule
\rowcolor{vihbg}
UserKNN & 0.129 & 0.064 & 0.119 & 0.052\\
SASRec & \textbf{0.168} & \textbf{0.084} & \textbf{0.129} & \textbf{0.059} \\
SSE-PT & \textbf{0.168} & 0.082 & 0.125 & 0.056\\
MostPop & 0.110 & 0.054 & 0.121 & 0.057\\
EASE & 0.120 & 0.060 & 0.096 & 0.043\\
BPR-MF & 0.098 & 0.049 & 0.069 & 0.032\\
ItemKNN & 0.087 & 0.042 & 0.042 & 0.016 \\
\bottomrule
\end{tabular}
\end{table}

Three readings follow. First, the strict keys are harder. Evaluated with the freeze models, UserKNN reaches 0.085 on the 481 strict keys but 0.196 on the 317 keys that the filter removes, and SASRec 0.125 against 0.243. The removed keys are single-token or colliding names, often with long, popularity-heavy histories, and they raise the freeze averages. Second, retraining on the strict pool raises UserKNN on those keys from 0.085 to 0.119, consistent with keys that pool several travellers no longer acting as neighbours, while SASRec stays near its freeze-model level (0.129 $\pm$ 0.010 over five seeds, against 0.125) with only 29\% of the freeze training interactions. Third, the sequential advantage does not survive: on the strict pool, SASRec and SSE-PT are not significantly different from UserKNN at any of the five seeds (smallest $p{=}0.058$), whereas on the freeze SASRec is significant at every seed. MostPop, UserKNN, SASRec, and SSE-PT form one statistically indistinguishable group on the strict pool, and ItemKNN, BPR-MF, and EASE have lower point estimates than UserKNN in both pools.

We therefore report the freeze as the public benchmark and the strict pool as its sensitivity bound. A method that claims to beat UserKNN on ViHoRec should report both pools, because a gain on the freeze can come from keys that merge several travellers. The strict pool does not recover persons; it removes the collisions that the released fields can prove. Two coarser UserKNN ablations on an earlier two-way construction (entity resolution on or off, random versus temporal holdout, and the min-$k$ sweep) are kept in Appendix~\ref{app:ablations}; they use a different split and are not comparable with the freeze.

\subsubsection{Sensitivity to entity-resolution errors}
\label{sec:er-sensitivity}

The released resolver misses one duplicate in the labelled sheet, the TTC pair (Section~\ref{lbl:er-pairs}). To measure how this false split propagates, we merge the pair into one hotel identifier and rebuild the three-way split with the same rule. The merge touches 28 interactions, but the three on the second listing all belong to reviewer keys with fewer than four interactions, so the rebuilt training, validation, and test files and the 535-hotel catalogue are identical to the freeze, and no number in Tables~\ref{tab:baseline}, \ref{tab:CFresult}, \ref{tab:reference}, \ref{tab:strict}, \ref{tab:coldstart}, \ref{tab:buckettests}, and~\ref{tab:seeds} changes. As a stricter sensitivity check, we also merge the Intourco pair that annotator~A had judged a duplicate. The two merges touch 36 interactions, shrink the catalogue to 534 hotels, and put an affected hotel in the history of 14 of the 798 evaluated keys; no test target changes. UserKNN Recall@10 stays at 0.129, ItemKNN and EASE are unchanged, and BPR-MF moves from 0.098 to 0.107 ($p{=}0.41$, paired Wilcoxon; Table~\ref{tab:baseline}), a change of the size produced by its random negative sampling. SASRec at seed 42 stays at 0.172 (Table~\ref{tab:seeds}). The known resolver error therefore does not move the reported benchmark.

\subsubsection{Effect of train-history length}
\label{sec:coldstart}

\begin{table*}[ht]
\centering
\caption{Recall@10 by train-history length on the public freeze (798 keys). Fixed is Hybrid-fixed ($\lambda{=}0.7$). Adaptive is the validation-tuned gate ($\alpha_{\min}{=}0.85$, $n_0{=}6$, $\tau{=}1$). Uncon.\ is the same gate with $\alpha_{\min}{=}0$. BPR-MF is the single run at seed 0 (77 of 798 hits, Recall@10 0.0965); Table~\ref{tab:baseline} instead reports the mean of seeds 0--2 (0.0977). Other columns are deterministic. Entries are point estimates on the full catalogue; bold marks the highest estimate in each bucket, but no per-bucket difference between Adaptive Hybrid and UserKNN is significant (Table~\ref{tab:buckettests}).}
\label{tab:coldstart}
\small
\setlength{\tabcolsep}{4pt}
\begin{tabular}{@{}l r ccccccc@{}}
\toprule
\textbf{History} & \textbf{\#Keys} & \textbf{MostPop} & \textbf{UserKNN} & \textbf{BPR-MF} & \textbf{Content} & \textbf{Fixed} & \textbf{Adaptive} & \textbf{Uncon.} \\
\midrule
2 (shortest) & 183 & 0.1148 & 0.1311 & 0.0765 & 0.0437 & 0.1257 & \textbf{0.1475} & 0.0437 \\
3--4         & 209 & 0.0813 & \textbf{0.1244} & 0.0574 & 0.0287 & 0.1053 & 0.1196 & 0.0431 \\
5--10        & 196 & 0.0969 & 0.1276 & 0.1378 & 0.0714 & \textbf{0.1633} & 0.1327 & 0.1531 \\
11+          & 210 & \textbf{0.1476} & 0.1333 & 0.1143 & 0.0762 & \textbf{0.1476} & 0.1333 & 0.1333 \\
\bottomrule
\end{tabular}
\end{table*}

\begin{figure}[ht]
    \centering
    \includegraphics[width=0.8\columnwidth]{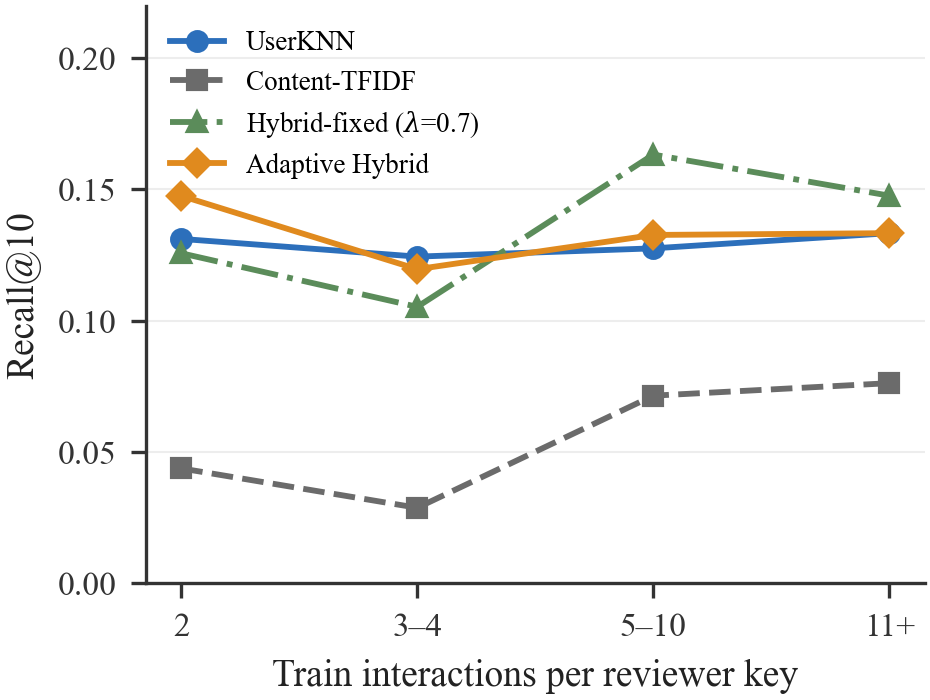}
    \caption{Recall@10 by train-history length for UserKNN, Content-TFIDF, Hybrid-fixed ($\lambda{=}0.7$), and Adaptive Hybrid on the public freeze.}
    \label{fig:adaptive-coldstart}
\end{figure}

Recall@10 depends on how many training interactions the reviewer key has (Table~\ref{tab:coldstart} and Figure~\ref{fig:adaptive-coldstart}). UserKNN stays between 0.124 and 0.133 across the four buckets. BPR-MF is lowest among the collaborative models at lengths 2 and 3--4 (0.077 and 0.057). Content-TFIDF stays below UserKNN in every bucket, including length 2 (0.044 vs.\ 0.131). Hybrid-fixed ($\lambda{=}0.7$) has the highest point estimate at lengths 5--10 (0.163) and ties MostPop at length $11+$ (0.148). Adaptive Hybrid has the highest point estimate only at length 2 (0.148 vs.\ UserKNN 0.131).

\begin{table}[ht]
\centering
\caption{Paired per-bucket tests on per-key Recall@10 (public freeze). Diff.: model minus UserKNN. Lo and Hi: bounds of the paired percentile bootstrap 95\% interval within the bucket (10{,}000 resamples). +/$-$: keys hit only by the model / only by UserKNN. $p_{\mathrm{Holm}}$: two-sided Wilcoxon, ties dropped, Holm-adjusted across the buckets that have discordant keys. $^{\dagger}$Unadjusted global test. The exact sign test gives the same conclusions. Shaded: the only bucket significant after Holm correction; bold: $p{<}0.05$.}
\label{tab:buckettests}
\small
\setlength{\tabcolsep}{2.5pt}
\begin{tabularx}{\columnwidth}{@{}>{\raggedright\arraybackslash}X r c c c c c@{}}
\toprule
\textbf{Hist} & \textbf{Keys} & \textbf{Diff.} & \textbf{Lo} & \textbf{Hi} & \textbf{+/$-$} & \textbf{$p_{\mathrm{Holm}}$} \\
\midrule
\multicolumn{7}{@{}l}{\textit{Adaptive Hybrid $-$ UserKNN}} \\
2      & 183 & $+0.016$ & $-0.016$ & $+0.049$ & 6/3 & 1.00\\
3--4   & 209 & $-0.005$ & $-0.024$ & $+0.014$ & 2/3 & 1.00\\
5--10  & 196 & $+0.005$ & $\phantom{+}0.000$ & $+0.015$ & 1/0 & 1.00\\
11+    & 210 & $\phantom{+}0.000$ & --- & --- & 0/0 & ---\\
All    & 798 & $+0.004$ & $-0.006$ & $+0.014$ & 9/6 & 0.44$^{\dagger}$ \\
\midrule
\multicolumn{7}{@{}l}{\textit{SASRec $-$ UserKNN}} \\
2      & 183 & $\phantom{+}0.000$ & $-0.049$ & $+0.049$ & 10/10 & 1.00\\
3--4   & 209 & $+0.053$ & $-0.005$ & $+0.110$ & 24/13 & 0.14 \\
\rowcolor{vihbg}
5--10  & 196 & $+0.061$ & $+0.020$ & $+0.107$ & 16/4 & \textbf{0.029} \\
11+    & 210 & $+0.052$ & $+0.010$ & $+0.095$ & 17/6 & 0.065\\
All    & 798 & $+0.043$ & $+0.019$ & $+0.068$ & 67/33 & \textbf{$<$0.001}$^{\dagger}$ \\
\bottomrule
\end{tabularx}
\end{table}

Table~\ref{tab:buckettests} tests those point estimates. Within each bucket we pair the two models on the same keys, compute a percentile bootstrap interval for the difference in per-key Recall@10, and run the Wilcoxon test of Section~\ref{sec:protocol}; Holm's procedure corrects across the buckets. At length 2, Adaptive Hybrid and UserKNN differ on only 9 of 183 keys (6 hits for Adaptive Hybrid only, 3 for UserKNN only). The difference is $+0.016$ with a 95\% interval of $[-0.016, +0.049]$ and a Holm-adjusted $p$ of 1.00. No bucket shows a significant difference, and at length $11+$ the two rankers make identical hits because the gate has returned almost all weight to UserKNN. The length-2 gap is therefore not evidence for history gating; with $\alpha_{\min}{=}0.85$ the gate changes the top~10 for few keys, and the data cannot distinguish the two rankers. For contrast, SASRec's advantage over UserKNN is zero at length 2 (0.131 vs.\ 0.131) and appears at longer histories: $+0.061$ at lengths 5--10 (Holm-adjusted $p{=}0.029$), with point gains of $+0.053$ at 3--4 and $+0.052$ at $11+$ that do not survive correction. In this split, sequence models gain where there is a sequence to model, and nothing tested here beats UserKNN on the shortest histories.

\subsubsection{Collaborative-filtering floor}
\label{sec:adaptive-analysis}

\begin{figure}[ht]
    \centering
    \includegraphics[width=0.8\columnwidth]{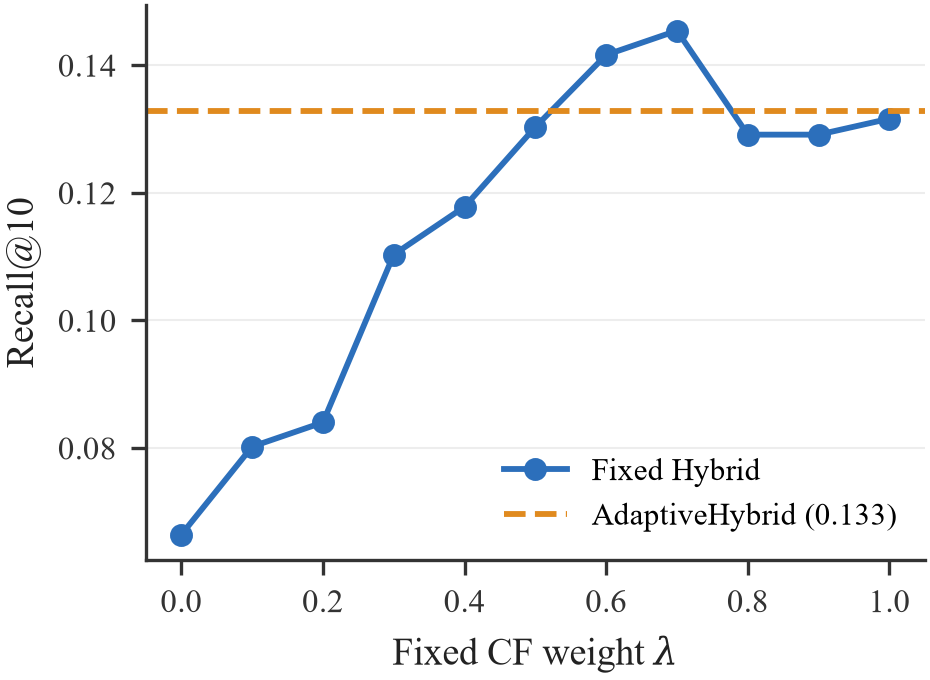}
    \caption{Fixed-$\lambda$ Recall@10 on validation. The dashed line is validation-tuned Adaptive Hybrid on the test fold ($0.133$); the best fixed $\lambda$ on test is $0.135$.}
    \label{fig:hybrid-lambda}
\end{figure}

No single fixed weight dominates validation (Figure~\ref{fig:hybrid-lambda}). On the test fold, validation-tuned Adaptive Hybrid reaches Recall@10 $= 0.133$, and the best fixed $\lambda$ reaches 0.135; neither differs significantly from UserKNN ($p{=}0.44$ and $p{=}0.50$). The Uncon.\ column of Table~\ref{tab:coldstart} removes the floor by setting $\alpha_{\min}{=}0$. Short histories then place nearly all of the weight on Content-TFIDF, and length-2 Recall@10 falls to 0.044, the level of content-only ranking. This is the clearest result about the metadata: on its own, at 53.5\% coverage, Content-TFIDF is far weaker than UserKNN at every history length, so any useful fusion must keep the collaborative score dominant. Validation selects $\alpha_{\min}{=}0.85$ for that reason, and with that floor the fused ranking differs from UserKNN only on a few keys (Table~\ref{tab:buckettests}). Adaptive Hybrid is therefore a diagnostic of the released metadata, not a recommended ranker. A learned or multilingual content encoder can replace $\alpha(n)$ on this split, with UserKNN as the floor to beat at short histories and SASRec as the sequential reference.

\section{Pseudonymization, Ethics, and License}
\label{sec:ethics}

\subsection{Pseudonymization, not anonymization}
The raw crawl contains reviewer display names. For the public release we \emph{remove} those names and replace them with a keyed hash $\text{HMAC-SHA256}(\text{salt}, \text{name})[{:}12]$. This is \emph{pseudonymization}: the same display name maps to the same identifier across versions, and a holder of the secret salt could re-identify keys. It is not anonymization in the sense that individuals cannot be re-identified. The salt is kept outside the repository (environment variable \texttt{VIHOREC\_SALT}); the name$\rightarrow$id map is never published. No emails, platform account IDs, or full review text are released. Location is city-level only. Because the name space is small, HMAC with a secret salt reduces casual dictionary attacks relative to a public hash, but reviewer-key collisions remain a validity threat (Section~\ref{sec:user-identity}).

\subsection{Ethical assessment}
Source platforms are Booking.com, Traveloka, and Ivivu. We collected only publicly visible ratings, dates, hotel names, and city locations---no private account pages. Direct identifiers in the release: none. Residual personal data: HMAC of a display name (reversible with the salt), city, and a numeric rating. Potential harm is mainly re-identification of prolific reviewers via distinctive rating sequences; we mitigate by withholding review text and the name map, and by labelling keys as reviewer names rather than people. Consent was not obtained from individual reviewers; the data were public at crawl time. Hotel-catalogue validation was labelled by co-authors and colleagues on public listings, not a human-subjects experiment. We did not run an institutional ethics-review protocol. A takedown request to the corresponding author removes rows from subsequent versions.

\subsection{Platform terms of service and legal residual}
\label{sec:tos}
Booking.com, Traveloka, and Ivivu restrict automated collection and commercial reuse of platform content. We hold no platform license and do not claim that redistribution of platform-derived fields is legally cleared. Residual legal risk remains. To keep a research-use position we: (i) collect only publicly visible scores and hotel metadata; (ii) do not redistribute raw HTML or full review text; (iii) release ratings, dates, hotel names and locations, derived metadata, and HMAC identifiers under CC BY-NC 4.0, and honour a takedown request to the corresponding author. What is \emph{not} released: raw HTML, full review text, display names, and the name$\rightarrow$id map. The authors are based in Vietnam. Users of the dataset must still comply with the source platforms' terms in their own jurisdiction.

\subsection{License and FAIR principles}
The ViHoRec dataset is released under the \emph{Creative Commons Attribution-NonCommercial 4.0 (CC BY-NC 4.0)} license; the pipeline code is released under the MIT license. Following the FAIR principles, each location has one role. The pipeline and benchmark code are on GitHub (\url{https://github.com/MinhNguyenDS/ViHoRec}); the processed dataset, split files, and result files are on Hugging Face (\url{https://huggingface.co/datasets/MinhDS/ViHoRec}). Both carry the version tag \texttt{v2.0} used for every number in this paper, and earlier tags stay available. The release includes a datasheet that documents provenance and schema, uses stable identifiers, and is distributed as UTF-8 CSV and Parquet.

\section{Conclusion}
\label{sec:conclusion}
\label{sec:limitations}

This paper releases ViHoRec, a Vietnamese hotel recommendation resource with city- and type-aware entity resolution, quantitative quality control, HMAC pseudonyms, and a short-history benchmark under one evaluation protocol. Its main contribution is construction evidence: the resolver's error rates carry exact intervals, its known errors do not move the benchmark, and the limits of the identifier are measured rather than assumed. On the public freeze, sequential self-attention ranks highest and UserKNN is the strongest closed-form method; on a stricter reviewer-key pool the sequential advantage is no longer significant, so both pools should be reported together. Adaptive Hybrid shows that, at 53.5\% coverage, the released metadata adds no measurable gain over UserKNN at any history length.

Several boundaries remain. Reviewer keys are display names, not persons: keys with a proven same-day multi-city collision carry at least 19.6\% of interactions, and the strict pool removes provable collisions without recovering persons. Chronology is enforced within each key rather than across the dataset, so the results describe per-key next-item ranking, not a prospective forecast. The resolver makes no observed false merge (exact 95\% upper bound 0.16) and one false split among 22 sampled duplicates, but errors outside the labelled sheet may remain. Text features cover 53.5\% of benchmark hotels, ratings are a single aggregate score, destinations are leisure-oriented, and 74 test targets are unreachable under the mask. Neural models other than SASRec and SSE-PT are single runs, and we hold no platform license (Section~\ref{sec:tos}). A global-cutoff companion split, review text where platform terms allow, and a person-level identity audit are natural next steps.

\section*{Acknowledgements}
We thank the user communities on the three platforms Booking.com, Traveloka, and Ivivu---the public data source that made the ViHoRec dataset possible. We thank Thuat Thien Nguyen and Minh Nhat Ta (University of Information Technology, VNU-HCM) for applying the ViHoRec manual validation guidelines to a stratified sample of the released corpus.
\bibliography{vihorec}
\bibliographystyle{acl_natbib}

\clearpage

\appendix
\captionsetup{hypcap=false}
\section{Entity-resolution labelling rules}
\label{app:llm-prompt}

The gold labels for entity resolution are human labels (Section~\ref{lbl:er-pairs}). Every annotator applies the written rules below to one pair at a time.

\subsection{Labelling rules}
The task is to decide whether the two listings are the same physical hotel. Answer $1$ when the names differ only by diacritics or transliteration (``Phú Quốc'' / ``Phu Quoc''), a property-type affix (``Khách sạn X'', ``X Hotel'', ``Khu nghỉ dưỡng X'', ``X Resort''), word order or punctuation, or an operator spelling of the same property (``Mövenpick Resort Phan Thiet'' / ``Khu nghỉ dưỡng Movenpick Phan Thiết''). Answer $0$ when the cities differ, even under one brand (``Raon Hotel'' in Đà Nẵng vs.\ Quy Nhơn); when tier or sub-brand markers differ (``Mường Thanh Grand'' vs.\ ``Mường Thanh Luxury''); when branch numbers differ (``La Cactus Hotel'' vs.\ ``La Cactus Hotel 2''); or when the property types genuinely differ (``Raon Hotel'' vs.\ ``Raon Villa''). A blank city is unknown, not a match. If the pair is still inconclusive, answer $0$. Judge each pair on its own, without balancing the number of $1$s and $0$s.

Each row of the labelling sheet names the two listings and nothing else:

\begin{quote}
\footnotesize
Listing A: name, city, platforms.\\
Listing B: name, city, platforms.\\
Do A and B refer to the same physical hotel?
\end{quote}

\subsection{Human labelling}
Two annotators apply the same $1$/$0$ rules to every one of the 426 pairs. Each works alone and sees only the two listings (name, city, platforms); the packet hides matcher decisions, the stratum, and why a pair was sampled, and rows are shuffled with a different seed per annotator. Annotators may consult public listings or maps and record the source in a notes column. Annotator~A labels the pairs in two blind sessions: 130 pairs and the remaining 296; annotator~B labels all 426. Pairs on which A and B disagree go to a third annotator~C, who did not build the resolver, applies the same rules, may consult public listings, and records the source and a confidence level; C's label is final. Agreement is reported as Cohen's $\kappa$ with a bootstrap interval, together with raw, positive, and negative agreement and PABAK, because the 22 duplicates make the classes highly unbalanced.

\section{Manual validation guidelines}
\label{app:guidelines}

This appendix reproduces the validation protocol (Section~\ref{lbl:manual}). Automated criteria C1--C3 are \emph{pass rates}, not human raters. Hotel-catalogue entries are labelled by three humans (the first author, Thuat Thien Nguyen, and Minh Nhat Ta). Entity-resolution pairs are labelled by two humans with adjudication (Appendix~\ref{app:llm-prompt}).

\subsection{Automated record checklist (C1--C3)}
\begin{enumerate}
    \item \textbf{C1 field validity:} rating in $[1, 10]$; date parseable and within 2010--2024; identifiers and source non-empty.
    \item \textbf{C2 release consistency:} the tuple exists in the released tables with matching rating, date, and source.
    \item \textbf{C3 identity plausibility:} reviewer name is not a placeholder and is longer than three characters; hotel location is not a multi-city conflict.
\end{enumerate}

\subsection{Location-consistency review}
After entity resolution, any \texttt{hotel ID} mapped to more than one city is flagged. Three annotators inspect the flagged listings independently, one platform each (Booking.com, Traveloka, Ivivu), without seeing one another's labels. If the identifier still joins two cities, it is split rather than released as one hotel.

\subsection{Hotel-catalogue human labels}
Three annotators independently mark each sampled hotel as a valid catalogue entry or not (non-empty name and city; unique \texttt{hotel\_id}; at least one interaction). Disagreements go to adjudication. Report percent agreement; do not treat C1--C3 as a three-rater $\kappa$.

\section{Dataset field reference and sample records}
\label{app:schema}
This appendix documents the concrete schema of each released file and provides representative records. All files are UTF-8 CSV unless noted otherwise.

\subsection{\texttt{interactions.csv} (17{,}911 rows)}
Each row is one user--hotel rating event.

\begin{center}
\small
\schematable{
\toprule
\textbf{Field} & \textbf{Type} & \textbf{Description} \\
\midrule
\texttt{user\_id} & string & Salted-HMAC pseudonym; no real name \\
\texttt{hotel\_id} & string & Canonical hotel id (e.g., H0287) \\
\texttt{rating} & float & Score in $[1,10]$ \\
\texttt{date} & date & Review date (YYYY-MM-DD) \\
\texttt{source} & categorical & booking, traveloka, or ivivu \\
\bottomrule
}
\captionof{table}{Schema of \texttt{interactions.csv}.}
\label{tab:app-interactions-schema}
\end{center}

\begin{center}
\footnotesize
\setlength{\tabcolsep}{3pt}
\resizebox{\columnwidth}{!}{%
\begin{tabular}{@{}lllll@{}}
\toprule
\textbf{user\_id} & \textbf{hotel\_id} & \textbf{rating} & \textbf{date} & \textbf{source} \\
\midrule
\texttt{Udd12fb44b382} & \texttt{H0287} & 6.3 & 2011-10-15 & ivivu \\
\texttt{U8253b44df1a6} & \texttt{H0392} & 6.7 & 2011-11-16 & ivivu \\
\texttt{U494dffb5df0b} & \texttt{H0409} & 7.3 & 2012-01-05 & ivivu \\
\bottomrule
\end{tabular}%
}
\captionof{table}{Sample records from \texttt{interactions.csv}.}
\label{tab:app-interactions-samples}
\end{center}

\subsection{\texttt{users.csv} (6{,}822 rows)}
One row per distinct pseudonymous reviewer key; activity count only.

\begin{center}
\small
\schematable{
\toprule
\textbf{Field} & \textbf{Type} & \textbf{Description} \\
\midrule
\texttt{user\_id} & string & Same pseudonym as in interactions.csv \\
\texttt{n\_interactions} & integer & Total ratings by this reviewer key in the corpus \\
\bottomrule
}
\captionof{table}{Schema of \texttt{users.csv}.}
\label{tab:app-users-schema}
\end{center}

\begin{center}
\small
\setlength{\tabcolsep}{4pt}
\begin{tabularx}{\columnwidth}{@{}S{2.5cm}r@{}}
\toprule
\textbf{user\_id} & \textbf{n\_interactions} \\
\midrule
\texttt{U82a5ed2b4568} & 210 \\
\texttt{U16c2298dc0c0} & 188 \\
\texttt{U6a1ec282d822} & 172 \\
\bottomrule
\end{tabularx}
\captionof{table}{Sample records from \texttt{users.csv} (most active reviewer keys after placeholder drop).}
\label{tab:app-users-samples}
\end{center}

\subsection{\texttt{hotels.csv} (560 rows)}
Canonical hotel catalogue after cross-site entity resolution.

\begin{center}
\small
\schematable{
\toprule
\textbf{Field} & \textbf{Type} & \textbf{Description} \\
\midrule
\texttt{hotel\_id} & string & Stable id (e.g., H0000) \\
\texttt{name} & string & Representative hotel name (Vietnamese/English) \\
\texttt{location} & string & City/region in Vietnam \\
\bottomrule
}
\captionof{table}{Schema of \texttt{hotels.csv}.}
\label{tab:app-hotels-schema}
\end{center}

\begin{center}
\footnotesize
\setlength{\tabcolsep}{3pt}
\begin{tabularx}{\columnwidth}{@{}l>{\raggedright\arraybackslash}X l@{}}
\toprule
\textbf{hotel\_id} & \textbf{name} & \textbf{location} \\
\midrule
\texttt{H0000} & Dragon King 1 Hotel Da Lat & Da Lat \\
\texttt{H0005} & Khanh Uyen 1 Hotel & Da Lat \\
\texttt{H0002} & Pho Hoi 1 Hotel & Hoi An \\
\bottomrule
\end{tabularx}
\captionof{table}{Sample records from \texttt{hotels.csv}.}
\label{tab:app-hotels-samples}
\end{center}

\subsection{Content metadata (\texttt{data\_content\_based\_raw.csv}, 309 rows)}
Rich hotel attributes for content-based recommendation; not part of the pseudonymized interaction release but shipped with the repository.

\begin{center}
\footnotesize
\setlength{\tabcolsep}{3pt}
\begin{tabularx}{\columnwidth}{@{}>{\raggedright\arraybackslash}p{2.8cm}>{\raggedright\arraybackslash}X@{}}
\toprule
\textbf{Field} & \textbf{Description} \\
\midrule
\texttt{Location} & City/region \\
\texttt{NameHotel} & Hotel name \\
\texttt{RatingHotel} & Aggregate score on the platform \\
\texttt{CountRating} & Number of platform reviews \\
\texttt{Price} & Typical price (VND) \\
\texttt{Facilities} & Comma-separated amenity tags (Vietnamese) \\
\texttt{Quality} & Star/category level (integer) \\
\texttt{DistanceCenter} & Distance to city centre (km) \\
\texttt{Around} & Nearby place types (Vietnamese) \\
\texttt{Vicinity} & Vicinity categories (Vietnamese) \\
\texttt{Link} & Booking URL \\
\bottomrule
\end{tabularx}
\captionof{table}{Schema of the content-metadata file.}
\label{tab:app-content-schema}
\end{center}

\begin{center}
\footnotesize
\setlength{\tabcolsep}{3pt}
\begin{tabularx}{\columnwidth}{@{}>{\raggedright\arraybackslash}p{2.2cm}>{\raggedright\arraybackslash}X >{\raggedright\arraybackslash}p{2.8cm}@{}}
\toprule
\textbf{Field} & \textbf{Example value} & \textbf{Note} \\
\midrule
\texttt{Location} & Hue & --- \\
\texttt{NameHotel} & The Sunriver Boutique Hotel Hue & --- \\
\texttt{Facilities} & Outdoor pool, free WiFi, \ldots & Structured tags, not a user review \\
\texttt{Around} & Bridge, Cafe/bar & --- \\
\texttt{Vicinity} & Heritage site, museum, train, airport & --- \\
\bottomrule
\end{tabularx}
\captionof{table}{Sample content-metadata fields (record: The Sunriver Boutique Hotel Hue, Hue).}
\label{tab:app-content-samples}
\end{center}

\begin{table*}[ht]
\centering
\caption{Per-seed results of SASRec and SSE-PT on the public freeze and on the strict reviewer-key pool, under the protocol of Table~\ref{tab:protocol}. $p$: two-sided paired Wilcoxon test against UserKNN on per-key Recall@10 for that run. Shaded: the mean over the five seeds, as reported in Tables~\ref{tab:CFresult} and~\ref{tab:strict}; SD: sample standard deviation over the same runs.}
\label{tab:seeds}
\small
\setlength{\tabcolsep}{5pt}
\begin{tabular}{@{}lcccccccc@{}}
\toprule
 & \multicolumn{4}{c}{\textbf{SASRec}} & \multicolumn{4}{c}{\textbf{SSE-PT}} \\
\cmidrule(lr){2-5}\cmidrule(lr){6-9}
\textbf{Seed} & \textbf{R@10} & \textbf{N@10} & \textbf{MRR} & $p$ & \textbf{R@10} & \textbf{N@10} & \textbf{MRR} & $p$ \\
\midrule
\multicolumn{9}{@{}l}{\textit{Public freeze (798 keys; UserKNN R@10 0.129)}}\\
42 & 0.1717 & 0.0845 & 0.0753 & $<$0.001 & 0.1717 & 0.0810 & 0.0706 & $<$0.001\\
43 & 0.1679 & 0.0827 & 0.0740 & $<$0.001 & 0.1754 & 0.0852 & 0.0745 & $<$0.001\\
44 & 0.1754 & 0.0868 & 0.0767 & $<$0.001 & 0.1767 & 0.0846 & 0.0729 & $<$0.001\\
45 & 0.1591 & 0.0802 & 0.0736 & 0.008 & 0.1516 & 0.0786 & 0.0737 & 0.055\\
46 & 0.1679 & 0.0855 & 0.0774 & $<$0.001 & 0.1629 & 0.0798 & 0.0716 & 0.004 \\
\rowcolor{vihbg}
Mean & 0.1684 & 0.0839 & 0.0754 & --- & 0.1677 & 0.0818 & 0.0727 & ---\\
SD   & 0.0061 & 0.0026 & 0.0017 & --- & 0.0105 & 0.0029 & 0.0016 & --- \\
\midrule
\multicolumn{9}{@{}l}{\textit{Strict pool (481 keys; UserKNN R@10 0.119)}}\\
42 & 0.1206 & 0.0577 & 0.0560 & 0.866 & 0.1289 & 0.0570 & 0.0515 & 0.411\\
43 & 0.1206 & 0.0559 & 0.0537 & 0.869 & 0.1227 & 0.0548 & 0.0509 & 0.739\\
44 & 0.1331 & 0.0643 & 0.0595 & 0.262 & 0.1206 & 0.0566 & 0.0538 & 0.879\\
45 & 0.1435 & 0.0628 & 0.0544 & 0.058 & 0.1227 & 0.0562 & 0.0522 & 0.785\\
46 & 0.1289 & 0.0545 & 0.0487 & 0.411 & 0.1310 & 0.0539 & 0.0471 & 0.317 \\
\rowcolor{vihbg}
Mean & 0.1293 & 0.0590 & 0.0545 & --- & 0.1252 & 0.0557 & 0.0511 & ---\\
SD   & 0.0096 & 0.0043 & 0.0039 & --- & 0.0045 & 0.0013 & 0.0025 & --- \\
\bottomrule
\end{tabular}
\end{table*}

\subsection{Benchmark split (\texttt{benchmark/})}
Public three-way temporal leave-one-out split for short-history evaluation (798 reviewer keys $\times$ 535 hotels). Per key, the last interaction is test, the second-last is validation, and the rest is train. Relevance is implicit next-item: all ratings are positives.

\begin{center}
\footnotesize
\setlength{\tabcolsep}{3pt}
\begin{tabularx}{\columnwidth}{@{}>{\raggedright\arraybackslash}p{2.6cm}>{\raggedright\arraybackslash}X >{\raggedright\arraybackslash}X@{}}
\toprule
\textbf{File} & \textbf{Fields} & \textbf{Role} \\
\midrule
\texttt{train.csv} & userID, itemID, rating, timestamp & Training interactions \\
\texttt{val.csv} & userID, itemID, rating, timestamp & Second-last interaction per user (tuning only) \\
\texttt{test.csv} & userID, itemID, rating, timestamp & Last interaction per user (final table only) \\
\texttt{user\_map.csv} & user\_id, userID & Reviewer key $\rightarrow$ integer id \\
\texttt{item\_map.csv} & hotel\_id, itemID & Hotel id $\rightarrow$ integer id \\
\texttt{split\_config.json} & protocol, relevance, counts & Split metadata \\
\texttt{PROTOCOL.md} & --- & Task name, folds, relevance definition \\
\bottomrule
\end{tabularx}
\captionof{table}{Benchmark auxiliary files.}
\label{tab:app-benchmark-schema}
\end{center}

\begin{center}
\footnotesize
\setlength{\tabcolsep}{5pt}
\begin{tabular*}{\columnwidth}{@{\extracolsep{\fill}}rrrr@{}}
\toprule
\textbf{userID} & \textbf{itemID} & \textbf{rating} & \textbf{timestamp} \\
\midrule
\multicolumn{4}{@{}l}{\textit{train.csv (first two rows)}} \\
0 & 27 & 7.3 & 1392336000 \\
0 & 355 & 9.0 & 1512432000 \\
\midrule
\multicolumn{4}{@{}l}{\textit{val.csv (first two rows)}} \\
0 & 99 & 7.0 & 1541721600 \\
1 & 473 & 9.0 & 1683158400 \\
\midrule
\multicolumn{4}{@{}l}{\textit{test.csv (first two rows)}} \\
0 & 99 & 8.0 & 1555459200 \\
1 & 472 & 6.0 & 1687478400 \\
\bottomrule
\end{tabular*}
\captionof{table}{Sample benchmark records (Unix seconds).}
\label{tab:app-benchmark-samples}
\end{center}

\section{Preliminary two-way ablations}
\label{app:ablations}

The effect of the resolver's known errors on the released benchmark is measured in Section~\ref{sec:er-sensitivity}. Tables~\ref{tab:ablation} and~\ref{tab:mink} keep two coarser UserKNN ablations for context. They use a different construction: a two-way split (800 keys, no validation fold) built before the placeholder rows were dropped, so their numbers are not comparable with the three-way freeze (UserKNN 0.129). On that construction, canonical hotel keys move UserKNN Recall@10 from 0.1288 (raw names) to 0.1338, and temporal leave-one-out gives 0.1338 against 0.1125 for a random holdout.

Table~\ref{tab:mink} varies the minimum interaction count on the same two-way construction. The public freeze sets $k{=}4$, which yields 798 keys at 97.60\% sparsity on the three-way matrix. At $k{=}8$, two-way Recall@10 is 0.154 on 364 keys. That cut describes a denser subpopulation.

\begin{table}[ht]
\centering
\caption{UserKNN ablations on the two-way split built before placeholder removal (800 keys). Not comparable with the three-way freeze.}
\label{tab:ablation}
\small
\setlength{\tabcolsep}{3pt}
\begin{tabularx}{\columnwidth}{@{}>{\raggedright\arraybackslash}X ccc@{}}
\toprule
\textbf{Setting} & \textbf{R@10}$\uparrow$ & \textbf{N@10}$\uparrow$ & \textbf{MRR}$\uparrow$ \\
\midrule
\multicolumn{4}{@{}l}{\textit{Entity resolution}}\\
\quad Off (raw hotel names) & 0.1288 & 0.0656 & 0.0623 \\
\rowcolor{vihbg}
\quad On (canonical)  & 0.1338 & 0.0671 & 0.0630 \\
\midrule
\multicolumn{4}{@{}l}{\textit{Evaluation protocol}}\\
\quad Random leave-one-out  & 0.1125 & 0.0566 & 0.0556 \\
\rowcolor{vihbg}
\quad Temporal       & 0.1338 & 0.0671 & 0.0630 \\
\bottomrule
\end{tabularx}
\end{table}

\begin{table}[ht]
\centering
\caption{Minimum-interaction threshold sweep (UserKNN, same two-way construction as Table~\ref{tab:ablation}). Three-way freeze at $k{=}4$ has 798 keys.}
\label{tab:mink}
\small
\setlength{\tabcolsep}{4pt}
\begin{tabular}{@{}rrrr@{}}
\toprule
\textbf{min-$k$} & \textbf{\#Keys} & \textbf{Sparsity (\%)} & \textbf{R@10}$\uparrow$ \\
\midrule
2 & 2{,}084 & 98.82 & 0.1248 \\
3 & 1{,}164 & 98.15 & 0.1349 \\
\rowcolor{vihbg}
4 (two-way) & 800 & 97.53 & 0.1338 \\
5 & 616 & 96.97 & 0.1347 \\
8 & 364 & 95.35 & 0.1538 \\
\bottomrule
\end{tabular}
\end{table}

\section{Seed repeats for the sequential models}
\label{app:seeds}

Tables~\ref{tab:CFresult} and~\ref{tab:strict} report SASRec and SSE-PT as means over five training runs; Table~\ref{tab:seeds} lists every run. Only the random seed (42--46) changes between runs. The configuration, the epoch selected on validation for each run, and the protocol of Table~\ref{tab:protocol} are fixed, and each run's $p$ is the two-sided paired Wilcoxon test against UserKNN on per-key Recall@10. On the public freeze, SASRec is significantly above UserKNN at every seed (largest $p{=}0.008$), and SSE-PT at four of five (seed 45: 0.152, $p{=}0.055$). On the strict pool, neither model differs significantly from UserKNN at any seed (smallest $p{=}0.058$). All other learned models are single runs at seed 42, and Random and BPR-MF in Table~\ref{tab:baseline} average seeds 0--2. The BPR-MF column of Table~\ref{tab:coldstart} is seed 0 only.

\end{document}